\documentclass[twocolumn]{aastex631}
\shorttitle{Pulse-Phase modulation in SGR1900;14}
\shortauthors{Makishima et al.}
\newcommand{\sgr}{SGR~1900+14}    
\newcommand{\bxp}{4U~1626$-$67}
\newcommand{\oneE}{1E~1547.0$-$5408}    
\newcommand{\NuS}{{\it NuSTAR}}
\newcommand{\Su}{{\it Suzaku}}
\usepackage{booktabs}
\shorttitle{Pulse-Phase modulation in SGR 1900+14}
\shortauthors{Makishima et al.}
\begin{document}
\received{2021/07/22}
\accepted{2021/09/20}

\title{Discovery of 40.5 ks Hard X-ray Pulse-Phase Modulations
from \sgr\ 
}
\author{K.  Makishima}
\affiliation{
Kavli Institute for the Physics and Mathematics of the Universe (WPI),\\
The University of Tokyo,
5-1-5 Kashiwa-no-ha, Kashiwa, Chiba 277-8683, Japan}
\affiliation{Department of Physics, The University of Tokyo,
7-3-1 Hongo, Bunkyo-ku, Tokyo 113-0033, Japan}
\affiliation{High Energy Astrophysics Laboratory, and MAXI Team, 
The Institute of Physical and \\Chemical Research (RIKEN),
2-1 Hirosawa, Wako, Saitama 351-0198, Japan}
\author{T. Tamba}
\affiliation{Department of Physics, The University of Tokyo,
7-3-1 Hongo, Bunkyo-ku, Tokyo 113-0033, Japan}
\author{Y. Aizawa}
\affiliation{Department of Physics, The University of Tokyo,
7-3-1 Hongo, Bunkyo-ku, Tokyo 113-0033, Japan}
\author{H. Odaka}
\affiliation{Department of Physics, The University of Tokyo,
7-3-1 Hongo, Bunkyo-ku, Tokyo 113-0033, Japan}
\affiliation{Research Center for the Early Universe, School of Science, \\
The University of Tokyo, 
7-3-1 Hongo, Bunkyo-ku, Tokyo 113-0033, Japan}
\author{H. Yoneda}
\affiliation{RIKEN Nishina Center, 2-1 Hirosawa, Wako, Saitama 351-0198, Japan}
\author{T. Enoto}
\affiliation{Extreme Natural Phenomena RIKEN Hakubi Research Team,\\
The Institute of Physical and Chemical Research,
2-1 Hirosawa, Wako, Saitama 351-0198, Japan}
\author{H. Suzuki}
\affiliation{Department of Physics, Konan University,
8-9-1 Okamoto, Higashi-Nada-ku, Kobe 658-8501, Japan}
\email{maxima@phys.s.u-tokyo.ac.jp}

\begin{abstract}
X-ray timing properties of  the magnetar \sgr\ were studied,
using the data taken with \Su\ in 2009  and \NuS\ in 2016,
for a  time lapse of 114 ks and 242 ks, respectively.
On both occasions, the object exhibited
the characteristic two-component spectrum.
The soft component, dominant in energies below $\sim 5$ keV,
showed a regular pulsation, 
with a period of $P=5.21006$ s  as determined with the \Su\ XIS,
and $P=5.22669$ with \NuS.
However, in $\gtrsim 6$ keV where  the hard component dominates,
the pulsation became  detectable  with the \Su\ HXD and  \NuS,
only after the data were corrected for periodic pulse-phase modulation,
with a period of $T=40-44$ ks and an amplitude of $\approx 1$ s.
Further correcting the two data sets  for complex 
energy dependences in the phase-modulation parameters,
the hard X-ray pulsation became fully detectable,
in 12--50 keV with the HXD, and 6--60 keV with \NuS,
using a common value of  $T=40.5 \pm 0.8$ ks.
Thus, \sgr\ becomes a third example, 
after 4U 0142+61 and 1E 1547$-$5408,
to show the hard X-ray pulse-phase modulation,
and a second case 
of  energy dependences in the modulation parameters.
The neutron star in this system is inferred to perform  free precession,
as it is axial deformed by $\approx P/T =1.3 \times 10^{-4}$
presumably due to
$\sim 10^{16}$ G toroidal magnetic fields.
As a counter example, the \Su\ data of the binary pulsar
\bxp\ were analyzed, but no similar effect was found.
These results altogether argue against the accretion scenario for magnetars.
\end{abstract}

\section{INTRODUCTION}
\label{sec:intro}
Through \Su\ and \NuS\  observations in hard X-rays of
the magnetars 4U 0142+61 \citep{Makishima14,Makishima19}
and \oneE\  (\citealt{Makishima16}; Makishima et al. 2021, hereafter Paper I),
we have found a novel timing phenomenon;
their hard X-ray pulses are phase-modulated
with a long period of 55 ks and 36 ks, respectively.
A likely origin of the effect \citep{Makishima14, Makishima16}
is that the neutron stars (NSs) in these systems harbor
toroidal magnetic fields $B_{\rm t}$ reaching $ \sim 10^{16}$ G,
and the magnetic pressure axially deforms the stars
by $\epsilon \equiv \Delta I/I \sim 10^{-4}$
where $I$ is the moment of inertia.
Then, the period $P_{\rm pr}$ of free-precession 
(= the pulse period) of the NS becomes slightly different  
from its rotation period $P_{\rm rot}$  around the symmetry axis,
as $P_{\rm pr} = (1+\epsilon) P_{\rm rot}$.
The beat between $P_{\rm pr}$  and $P_{\rm rot}$ 
appears at  the so-called  slip period given as 
\begin{equation}
T = P_{\rm pr} /\epsilon \cos \alpha~,
\label{eq:slip_period}
\end{equation}
where $\alpha$ is the wobbling angle
between the NS's symmetry axis 
and the angular momentum vector ${\bf L}$,
which are  fixed to the NS and the inertial frame, respectively.
We identify this $T$ with the observed pulse-phase-modulation periods.
Further studies of this phenomenon will provide 
valuable information on $B_{\rm t}$ of magnetars,
which are otherwise difficult to observationally estimate.

Of the two objects, 4U~0142+61 is an old magnetar
with  a characteristic age of $\tau_{\rm c}=65$ kyr,
a pulse period of  $P_{\rm pr}=8.69$ s,
and a rather stable X-ray intensity.
It represents a magnetar's subclass called 
Anomalous X-ray Pulsars.
In contrast, \oneE\  is a young object with $\tau_{\rm c}=0.7$ kyr
and the fastest rotation ($P_{\rm pr}=2.07$ s) 
among the confirmed magnetars, 
exhibiting  intensity
changes by 4 orders of magnitude \citep{Enoto17}.
The detection of the hard X-ray pulse-phase modulation
from these two contrasting magnetars suggests
that it  is a rather common phenomenon,
to be detected possibly from almost all objects of this class.

Among the past observations,
that of  \oneE\  with  \NuS\  is of particular interest,
because it allowed the discovery (Paper I)
of peculiar energy dependences
in the  pulse-phase modulation parameters.
Since this finding could provide valuable clues 
to the hard X-ray emission mechanism from magnetars,
we need to analyze the data of 
other magnetars for similar phenomena.
This makes  a second aim of the present study.

Our study has yet another aim.
The complex energy dependence in  \oneE\  found with \NuS\ (Paper I)
was not observed in the  \Su\ data of the same object
in an outburst \citep{Makishima16}.
Likewise, the modulation amplitude of  4U~0142+61 
derived with \NuS\  was much smaller than those 
measure with \Su\  on two occasions \citep{Makishima19}.
Thus, the \Su\ and \NuS\ results,
though generally consistent, are still subject to some differences,
or possibly discrepancies.
If  these two X-ray observatories give 
more consistent results on some other magnetars,
we will become more confident that we are not observing some
instrument-specific artifacts.

The above three aims urge us to perform 
detailed hard X-ray timing studies of other magnetars.
Obvious targets would be Soft Gamma Repeaters,
namely, another major subclass of magnetars.
In the present work, we hence select \sgr,
a  prototypical  objects of this subclass.
It  has $P_{\rm pr}=5.2$ s and $\tau_{\rm c}=0.9$ ks,
with  an estimated distance of $12.5$ kpc \citep{SGRdistance},
and exhibited a Giant Flare on 1998 August 27 (e.g., \citealt{Feroci01}).
Since it  was observed by both \Su\ and \NuS,
we utilize these archival data.

\section{OBSERVATIONS}

\subsection{\Su}
Throughout the 10 years of mission lifetime of \Su,
\sgr\ was observed twice.
One was a Target-of-Opportunity observation  (ObsID 401022010),
from 2006 April 01 UT 08:42:57 for a gross pointing of 47 ks
\citep{Nakagawa09,Enoto10}.
The other (ObsID 404077010) was
from 2009 April 26 UT18:23:44 for a gross pointing of 114 ks \citep{Enoto17}.
Since  the former would be too short,
we utilize the latter data set.
It was already used by \citet{Enoto17} in a summary study 
of the \Su\ observations of magnetars,
but detailed timing studies have not yet been conducted
at least to our knowledge.

In the 2009 observation,
the X-ray Imaging Spectrometer (XIS) 
covering 0.5--10 keV \citep{XIS}
was operated in the  1/4 window mode, 
with a time resolution of 2 s
which is somehow usable for the study of the 5.2 s  pulsation
\citep[see][]{Makishima14}.
From the three XIS cameras, XIS0, XIS1, and XIS3
that were operational at that time,
we extracted events using an accumulation radius of $2'$
around the source centroid.
When the three cameras are added up,
the source was detected with a count rate of
$0.36$ ct s$^{-1}$
(background subtracted but vignetting uncorrected).

In the same observation,
the Hard X-ray Detector (HXD; \citealt{HXD1,HXD2}) 
onboard \Su\ was operated in the standard mode.
We utilize the data from HXD-PIN,
with a nominal energy range of 10--70 keV,
where the source was detected with a 12--50 keV rate of
$0.046$ ct s$^{-1}$ after subtracting the modeled background
 \citep{Enoto17}.
The data from HXD-GSO are not utilized,
because the source was not detected significantly.

The arrival times of the XIS and HXD data
have been corrected for the barycentric motion of the Earth
and the spacecraft.

\subsection{\NuS}

From  2016 October 20 UT 16:56:08,
SGR1900+14 was observed with \NuS\
for an elapsed time of 242 ks (ObsID 30201013002).
The archival data were
already analyzed by \citet{Tamba19}, 
incorporating  {\it XMM-Newton} data which were partially simultaneous.
We do not utilize the  {\it XMM-Newton} data set,
because it covers only up to 10 keV
with a short exposure (23 ks).

As described in \citet{Tamba19}, 
the \NuS\ data were processed using 
{\tt nupipeline} and {\tt nuproducts} in HEASoft 6.23,
and the photon arrival times were corrected for the
Earth and spacecraft motions around the Solar system barycenter.
The net exposure was 122.6 ks 
after the standard pipeline processes.
Since the nearby bright binary GRS 1915+105 caused
severe stray light, particularly in FPMB,
we utilize only the FPMA data, 
like in \citet{Tamba19}.
To suppress the stray light,
the on-source FPMA events 
were extracted from a circular region of $R_{\rm acc}=50''$ radius
around the image centroid.
The source was detected with a background-subtracted
but vignetting-uncorrected count rate of
$0.055$ ct s$^{-1}$ in 3--70 keV.

\section{BASIC DATA ANALYSIS}
\label{sec:basic_ana}

\subsection{Light curves}
\label{subsec:light curves}

\begin{figure}[bth]
\centerline{
\includegraphics[width=8cm, height=3.5cm]{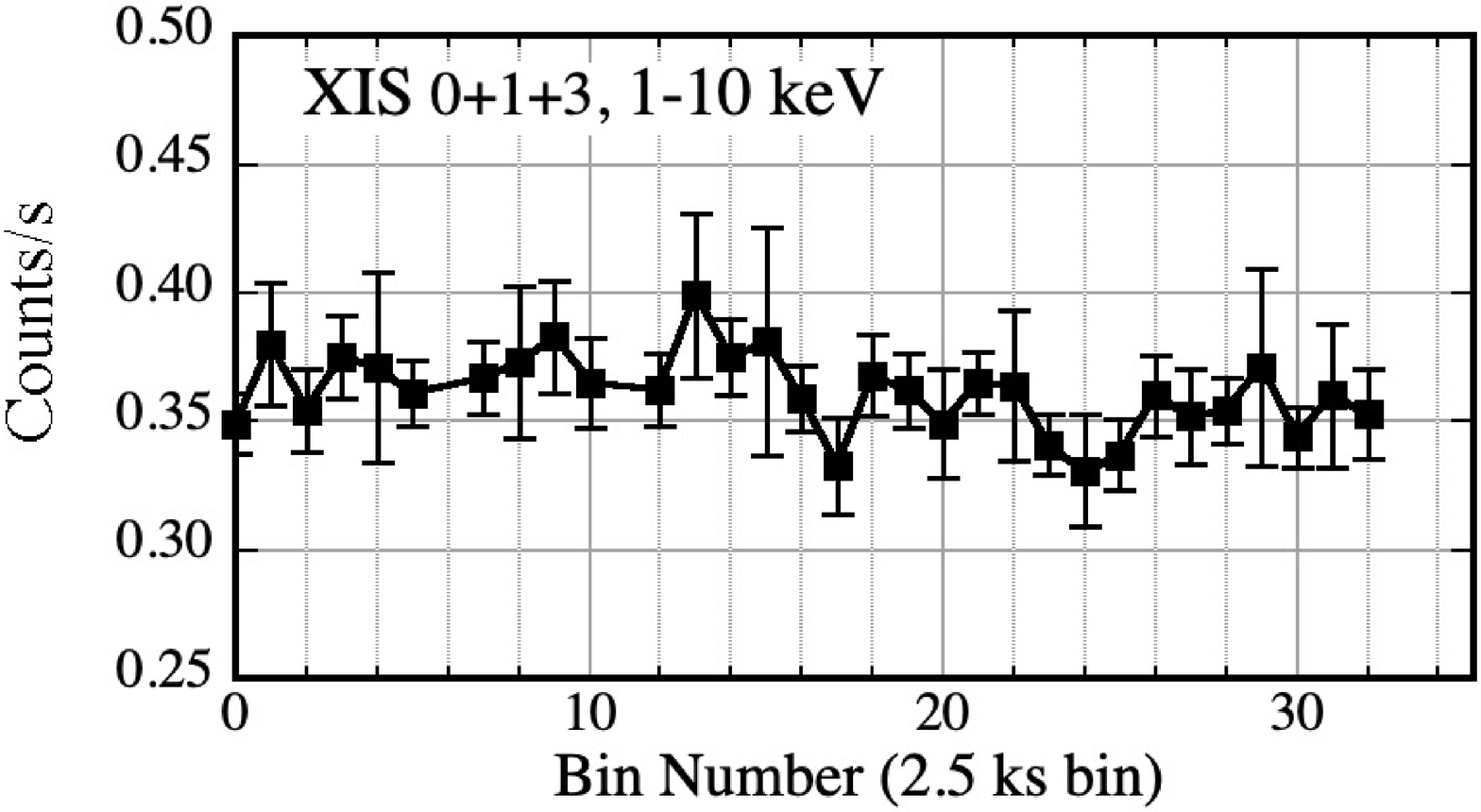}
}
\caption{Light curve of \sgr\ in 1--10 keV obtained with the \Su\ XIS
(the 3 cameras added together), with 2.5 ks binning.
The data include background (though negligible),
and are not corrected for the vignetting.
The data after bin 33 (82.5 ks) are omitted,
because of rather frequent data gaps.
}
\label{fig1:LC}
\end{figure}

Figure~\ref{fig1:LC} shows a 1--10 keV light curve of \sgr\
obtained with the \Su\ XIS.
The data, though suggesting some mild variations,
are statistically consistent with being constant.
We do not show light curves from the HXD,
because they are background-dominated, and  hence not informative.
Similarly we skip showing the \NuS\ light curve,
because it is already given in Fig.~5 of \citet{Tamba19}.

\subsection{Spectra}
\label{subsec:spectra}
\begin{figure}[hbt]
\centerline{
\includegraphics[width=8.2cm]{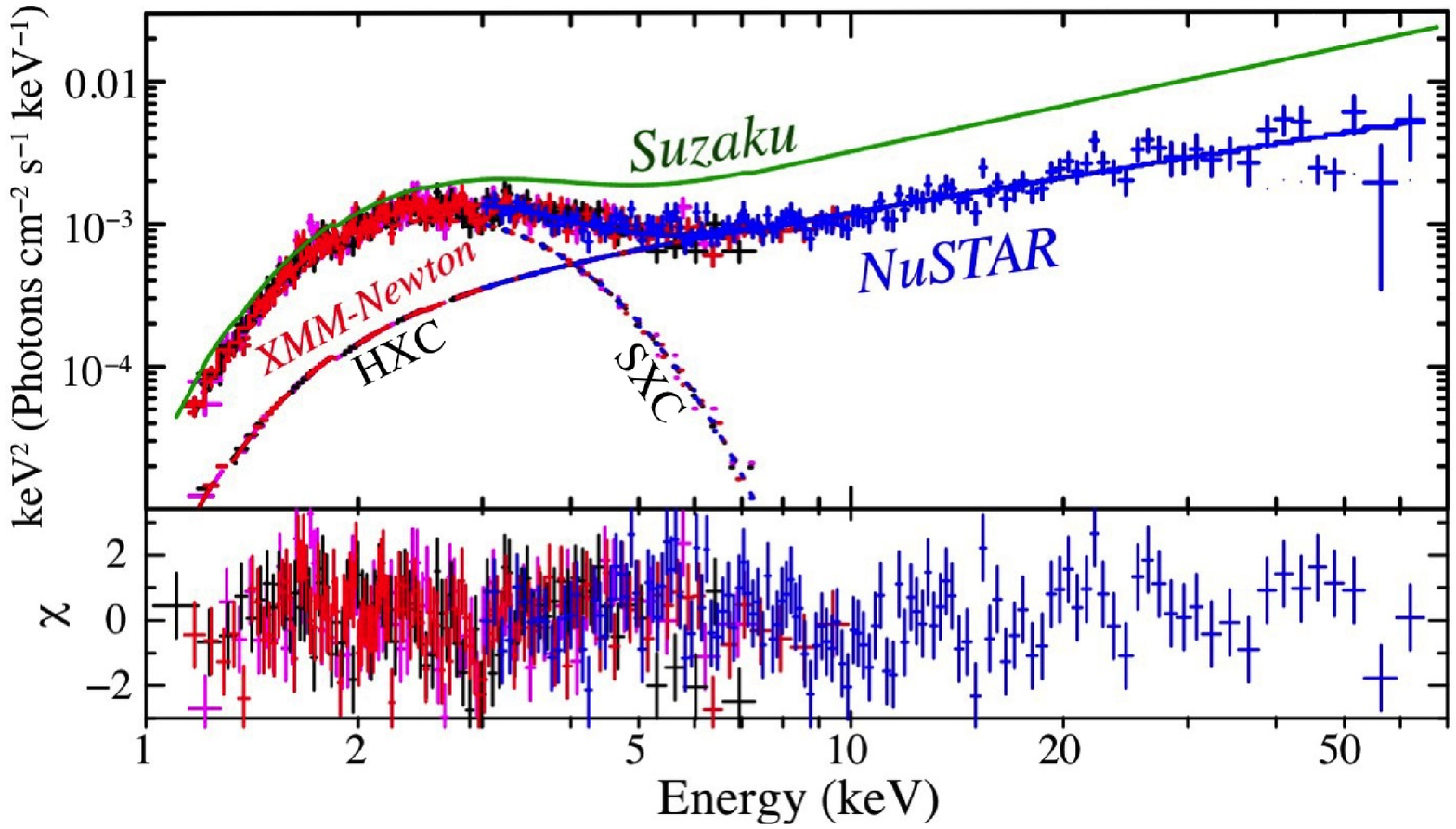}
}
\caption{
Background-subtracted  \NuS\  spectrum of \sgr\ in blue.
It is the same as Fig.\,4 of   \citet{Tamba19},
but converted to the $\nu F\nu$ form.
Partially simultatneous  {\it XMM-Newton} spectra
are shown in red (PN), magenta (MOS1), and  black (MOS2).
The \NuS\ and {\it XMM-Newton} spectra are fitted jointly,
by the sum of an absorbed blackbody for the SXC
and a power-law for the HXC.
The \Su\ spectrum \citep{Enoto17} is superposed as a green solid line,
in the best-fit model form jointly determined with the XIS and HXD.
}
\label{fig2:spectra}
\end{figure}

Figure~\ref{fig2:spectra} presents the background-subtracted \NuS\ 
FPMA+FPMB spectrum of \sgr,
and  partly simultaneous {\it XMM-Newton} spectra,
fitted jointly \citep{Tamba19} with a blackbody model for the Soft X-ray Component (SXC),
and a power-law model for the Hard X-ray Component (HXC).
The \Su\ spectrum, determined with the XIS and HXD \citep{Enoto17},
is superposed as a best-fit incident model.
As common among  magnetars \citep{Kuiper06,Enoto10},
the spectra on the two occasions  
consist of the SXC and the HXC
that  cross over at $\sim 5$ keV.
At  the 12.5 kpc distance,
the 1--60 keV luminosities measured with \Su\ and \NuS\ are
$5.5 \times 10^{35}$ erg s$^{-1}$ and $2.4\times 10^{35}$ erg s$^{-1}$, respectively.
These are typical of this object,
when it is not in an enhanced activity.

\subsection{Periodograms}
\label{subsec:PG_nodem}
To study the  pulsation expected  at a period of $P \approx 5.2$ s,
we  utilize so-called $Z_m^2$ periodograms,
wherein we epoch-fold the background-inclusive data 
into $N$ bins assuming a range of  trial periods,
and evaluate, at each $P$, the folded profile using $Z_m^2$ statistics
\citep{Z2_94, Makishima16, Makishima19}.
We do not correct each folded pulse profile for exposure,
because it is very uniform (to within 2\%) across the pulse cycle.
The profiles hence preserve the photon counts.

The quantity $Z_m^2$ is obtained by
summing up the Fourier power of the folded $N$-bin profile
up to a  specified  maximum harmonic number {$m$ ($\ll N$),
and normalizing the result to the total event number (Paper I).
The derived $Z_m^2$ is evaluated against 
$\chi^2$ distribution of $2m$ degrees of freedom
(with a mean of $2m$ and 1$\sigma$ scatter by $\sqrt{4m}$),
which  $Z_m^2$ would follow
if the data were dominated by Poisson noise.
As $m$ is increased, 
$Z_m^2$ also gets larger, but becomes more  noise dominated,
because the pulse signal is usually limited to lower harmonics like $m \leq 5$.  
For $m \rightarrow N$,
$Z_m^2$ approaches the usual chi-square of the pulse profile.
Therefore, the $Z_m^2$ method with a small $m$ is more noise tolerant
than the conventional chi-square  technique,
and unaffected by the choice of $N$
because  $Z_m^2$ is independent of  $N$ for $m \ll N$.

Figure \ref{fig3:PG_Suzaku_nodem}\,(a) shows a
1--10 keV pulse periodogram thus derived from  the \Su\ XIS
(the three cameras co-added).
As a pilot study,  we here employ  $m=2$,
considering that pulse profiles of some magnetars are  double peaked.
In spite of the limited  XIS time resolution (2.0 s in this case),
an outstanding peak with $Z_2^2=108$ has been revealed at a period of
\begin{equation}
P_{\rm XIS} =5.210\;06 \pm 0.000\;15~,
\label{eq:P0_XIS}
\end{equation}
where we estimated  the error  conservatively 
as the  half-width at half-maximum of the peak.
The two side lobes seen at $P=5.205$ s and $P=5.215$ s
are beat periods between $P_{\rm XIS}$ and the \Su's orbital period, 5.6 ks.
The three XIS cameras, when analyzed separately,
gave  consistent results, each with $Z_2^2$ from 26 to 42.

A hard X-ray periodogram from the same observation,
derived with the 12--50 keV \Su\ HXD-PIN data,
is presented in Fig.~\ref{fig3:PG_Suzaku_nodem}\,(b).
From the nominal energy range (10--70 keV) of HXD-PIN,
those below 12 keV and above 50 keV were discarded,
because of the dominance of thermal noise and particle background, respectively.
The inset shows a detail near  $P_{\rm XIS}$ of Equation~(\ref{eq:P0_XIS}).
Although we observe several peaks with $Z_2^2$ exceeding $\sim 15$
up to $\sim 20$,
none of them is dominating, 
and no enhancement is seen at $P_{\rm XIS}$, either.

\begin{figure}[t]
\centerline{
\includegraphics[width=7.5cm]{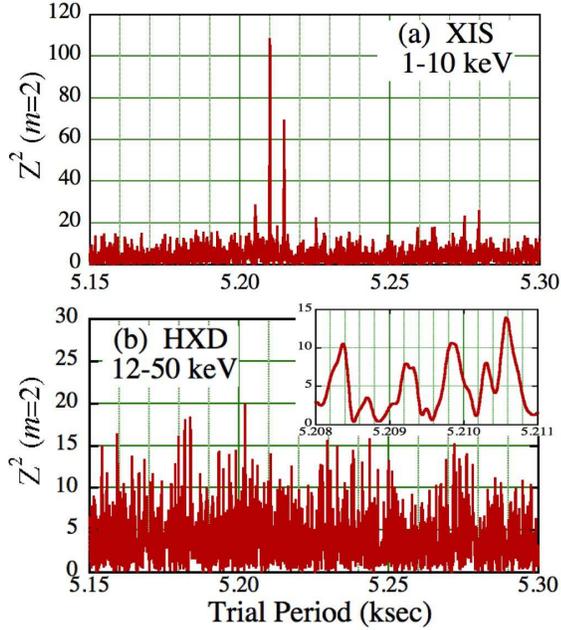}
}
\caption{Pulse periodograms of \sgr\ derived with \Su,
using the $Z_2^2$ statistics. Panel (a) is from the XIS in 1--10 keV,
whereas (b) is from the HXD in 12--50 keV.
The inset to (b)  gives a detail  near $P_{\rm XIS}$.
}
\label{fig3:PG_Suzaku_nodem}
\end{figure}

Similarly, we analyzed the \NuS\  FPMA data for the pulsation.
Figure~\ref{fig4:PG_NuSTAR_nodem} shows periodograms derived
in three typical energy bands.
Panel (a), using the 3--70 keV range,
reveals a clear peak reaching $Z_2^2=41.56$,
at a period of 
\begin{equation}
P_{\rm NuS} =5.226\;69 \pm 0.000\;03~,
\label{eq:P0_NuSTAR}
\end{equation}
which fully agrees  with  that reported in \citet{Tamba19}.
For reference, the probability of finding a 
value of $Z_2^2 \geq 41.56$ solely by chance is $2.1 \times 10^{-8}$.
It  still gives a very low post-trial chance probability of 
$Q =2.8 \times 10^{-5}$, after multiplied by 1330 
which is the Fourier wave number (the effective number of frequency trials)
contained in the period range of  
 Fig.~\ref{fig3:PG_Suzaku_nodem} and Fig.~\ref{fig4:PG_NuSTAR_nodem}.
We again observe two side lobes at $P=5.222$ s  and $P=5.232$ s,
arising from the beat with the {\it NuSTAR}'s orbital period, 5.8 ks.
When the  accumulation radius $R_{\rm acc}$  is  varied from $38''$ to $100''$,
the pulse significance became maximum
for $R_{\rm acc} \approx 50''$ which we have employed.

\begin{figure}[b]
\centerline{
\includegraphics[width=7.4cm]{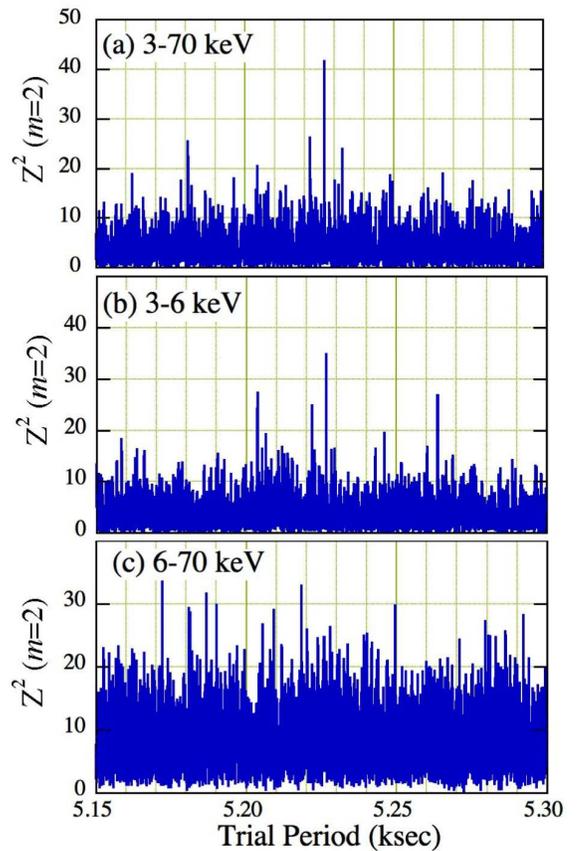}
}
\caption
{Same $Z_2^2$ periodograms as in Fig.~\ref{fig3:PG_Suzaku_nodem},
but derived with the \NuS\ FPMA in three energy ranges.
}
\label{fig4:PG_NuSTAR_nodem}
\end{figure}
The periodogram peak at $P_{\rm NuS}$ is 
reconfirmed with $Z_2^2=34.82$ also in panel (b),
which employs the 3--6 keV photons arising mainly from the SXC.
However, the 6--70 keV periodogram in panel (c),
representing the HXC,
does not show any dominant peak,
either at $P_{\rm NuS}$ or at any other period studied here.
Even when different energy intervals above 6 keV were employed,
the results did not change significantly.

In the SXC-dominant energies, 
thus the source pulsation has been detected significantly 
on  the two occasions,
and the derived periods are both consistent with the long-term
spin-down history of the source 
after the Giant Flare in 1998; see Fig. 11 of  \cite{Tamba19}.
In contrast, neither data set gave evidence of pulsation
in energies where the source signals are dominated by the HXC.
These results remain  unchanged
even if using higher values of $m$.
We present the folded pulse profiles of the SXC and HXC 
later in \S~\ref{subsec:Pr_nodem_dem}.

\section{DEMODULATION ANALYSIS}
\label{sec:demod}

\subsection{Demodulation formalism}
\label{subsec:formalism}
The apparent absence of the HXC pulsation, both in the \Su\ and \NuS\ data,
is reminiscent of the previous \Su\ results (\S~1)
on 4U~0142+61 in 2009 \citep{Makishima14}
and \oneE\ in the 2009 outburst \citep{Makishima16}.
In  these cases, the hard X-ray ($\gtrsim 10$ keV) pulsations
became detectable with high significance 
only after we correct, via so-called  demodulation procedure,
the photon arrival times for the pulse-phase modulation.
The same correction also  increased the 
HXC pulse significance in the \NuS\ data of \oneE\ 
in quiescence (Paper I).
Supposing that the HXC pulses of \sgr\ 
are in similar conditions,
we  apply the same timing corrections 
to the present two data sets.

The demodulation analysis assumes
that the arrival times $t$ of individual pulses  from the source 
are advancing/delaying periodically,  
by an amount
\begin{equation}
\delta t = A \sin (2\pi t/T -\psi)~,
\label{eq:modulation}
\end{equation}
compared to the case of an exactly regular pulsation.
Here, $T >0$, $A\geq 0$, and  $\psi$ ($0 \leq \psi < 2\pi$)
are  the period,  amplitude, and initial phase
of the assumed modulation, respectively.
Among them, $\psi$ can take any value between 0 and $2\pi$,
as it simply reflects when the data acquisition happened to start.
Then, by changing the arrival times of individual photons
(instead of pulses) from $t$ to $t -\delta t$,
we search for a set of parameters $(T, A, \psi)$ 
that maximize $Z_{\rm m}^2$ for the expected pulse period.

Although  $T$ is unknown,
a possible hint is provided by the inset to Fig.~\ref{fig3:PG_Suzaku_nodem}\,(b).
There, the periodogram shows several peaks 
separated in period by $\Delta P = (6-8) \times 10^{-4}$ s.
Such structures  could arise
if the main periodicity is modulated in its amplitude or phase,
at a long period of $P_0^2/\Delta P = (34-45)$ ks \citep{Makishima16},
where $P_0$ stands for $P_{\rm XIS}$ or $P_{\rm NuS}$.
Therefore, we set the search range of $T$
from 10 ks to 100 ks; 
at $T \lesssim 10$ ks, the analysis is often 
affected by the observatory' s orbital period,
and values of $T \gtrsim 100$ ks are not practical 
considering the overall data length (particularly of \Su).

\begin{figure*}[htb]
\centerline{
\includegraphics[width=14.3cm]{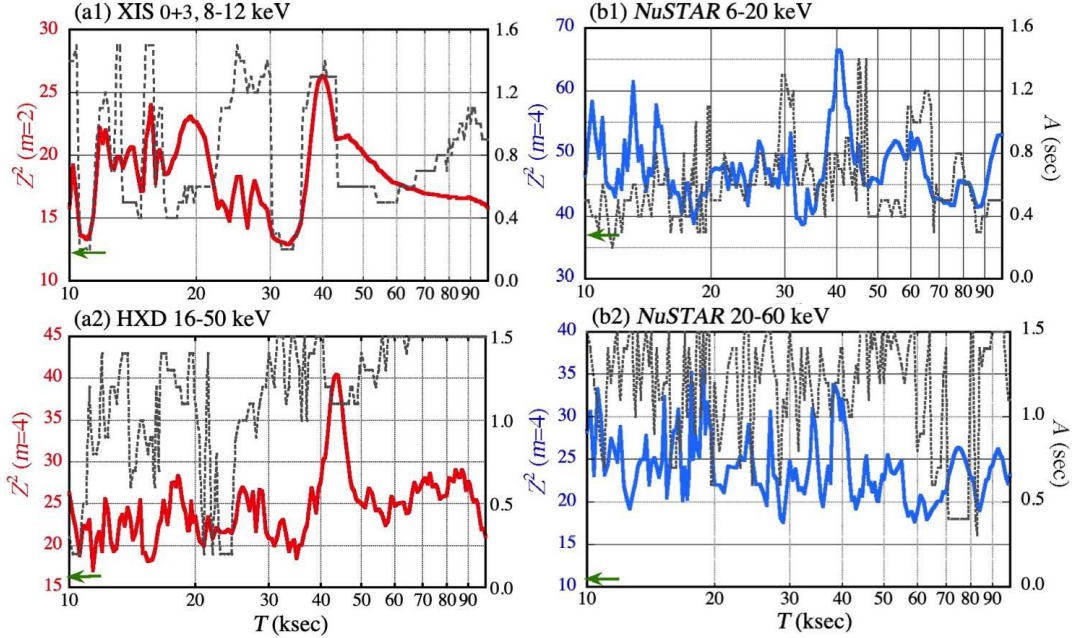}
}
\caption{DeMDs derived from the \Su\ XIS (cameras 0+3) 
in 8--12 keV with $m=2$ (panel a1),
the \Su\ HXD in 16--50 keV with $m=4$ (panel a2),
and from the 6--20 and 20--60  keV  \NuS\ data with $m=4$
(b1 and b2, respectively),
all including  background.
The abscissa is $T$ in ks,
and the red/blue curve (the left ordinate)
gives the maximum $Z_m^2$  obtained at each $T$
when $A$, $\psi$, and $P$ are all varied.
The dashed gray curve shows the value of $A$ (right ordinate)
that maximizes $Z_m^2$.
The green arrow at the bottom left of each panel
indicates $Z_m^2$  before the demodulation,
at $P \approx P_{\rm XIS}$ or $P \approx P_{\rm NuS}$.
}
\label{fig5:Tscan_4results}
\end{figure*}

\subsection{Demodulation diagrams (DeMDs)
\label{subsec:DeMDs}}

\subsubsection{\Su\ XIS data}
\label{subsubsec:demod_XIS}

The demodulation analysis was first applied  to the \Su\ XIS data,
using the 8--12 keV range where the HXC dominates.
Although the XIS energy response is poorly calibrated at $\gtrsim 10$ keV,
we included the 10--12 keV interval 
because  it still contains usable HXC signals,
and the calibration uncertainties does not affect timing studies.
We used the data from XIS0 and XIS3
(front-illuminated CCDs),
but not those of the XIS1 camera (back-illuminated CCD chips),
because its background at $\approx 10$ keV is
more than an order of magnitude higher 
than those of the other two cameras \citep{XIS}.
The maximum harmonic number of  $Z_m^2$ was tentatively set at $m=2$,
because any Fourier component  with $m \gtrsim 3$ of the  pulse profile
would be strongly smeared out by the XIS  time resolution.
As above, $T$ was varied over the 10--100 ks range,
with a step of 0.2 s to 1.0 s (depending on  $T$).
At each $T$, we scanned $A$ from 0 s to 1.5 s with a step of 0.1 s, 
$\psi$ from $0^\circ$  to $360^\circ$ with a $5^\circ$ step,
and $P$ over the error range of Equation~(\ref{eq:P0_XIS})
with a step of $20~\mu$s.

Figure~\ref{fig5:Tscan_4results}\,(a1) presents the maximum value of  $Z_2^2$ 
achieved at each $T$, when $A$, $\psi$, and $P$ are all optimized.
After Paper~I, 
this kind of plot is hereafter called {\it a demodulation diagram} (DeMD).
The result reveals a clear peak  at 
\begin{equation}
T = 40.0^{+3.3}_{-2.5}~~{\rm ks}~,
\label{eq:T_XIS}
\end{equation}
where the error is estimated as the range where $Z_m^2$ decreases
by 4.72 from the peak value (Paper I).
As shown in the same figure with a dashed gray curve,
this peak has  $A \approx 1.3$ s, or $\approx \pm 25\%$ of $P$.
For reference, the XIS0 and XIS3 data, when analyzed separately,
consistently reveal the 40 ks peak.
When the XIS1 data are included,
the peak becomes somewhat higher,
but the DeMD becomes  noisier in the $T=10-20$ ks interval,
presumably due to the  background variations.

As indicated by a green arrow in Fig.~\ref{fig5:Tscan_4results}\,(a1),
the XIS data give  $Z_2^2=12.29$  before the demodulation.
Therefore, the 40 ks peak in the DeMD, with $Z_2^2=26.36$,
yields  $\delta Z_2^2 = 14.07$,
where $\delta Z_m^2$ denotes a relative increment in $Z_m^2$,
and provides a measure of the pulse-significance increase
owing to the demodulation  (Paper I).
As a fiducial value, an increment by $\delta Z_m^2=10$
means a decrease in the  probability $Q$
by a factor of $0.67 \times 10^{-2} (1+\delta Z_m^2/Z_m^2)^{m-1}$,
or approximately by two orders of magnitude.

In Fig.~\ref{fig6:PG_nodem_dem}\,(a)
we compare two periodograms,
both computed using the 8--12 keV XIS 0+3 data.
The black one is before the demodulation,
whereas the red one is  after the demodulation 
employing Equation~(\ref{eq:modulation}) and $T=40.0$ ks,
together with  $A$ and $\psi$ as given in the figure.
The result visualizes 
that the demodulation selectively enhances $Z_2^2$
at $P \approx P_{\rm XIS}$, 
although it is not necessarily obvious
whether this peak is statistically significant.

\subsubsection{\Su\ HXD data}
\label{subsubsec:demod_HXD}

To the  background-inclusive HXD-PIN data,
we applied the  demodulation analysis 
with the same procedure as for the XIS, 
except  that the search step in $\psi$ is reduced to $3^\circ$
and  $m=4$ is employed.
The latter is because the HXC pulse profiles of 4U~0142+61 and \oneE,
though variable, often exhibit three to four peaks per cycle when demodulated,
and hence $m=4$ has generally been found most appropriate 
(\citealt{Makishima19}; Paper I).
In addition, we tentatively chose an energy range of 16--50 keV.
The derived DeMD is presented in Fig.~\ref{fig5:Tscan_4results}\,(a2),
where a prominent peak with $\delta Z_4^2 = 24.07$ is found at
\begin{equation}
T = 43.5 \pm 1.8~~{\rm ks}.
\label{eq:T_HXD}
\end{equation}
This $T$ is close to  Equation~(\ref{eq:T_XIS}),
and the best-estimated  amplitude of $A \approx 1.2$ s 
is consistent between the XIS and the HXD.
The parameters characterizing this DeMD peak are summarized in 
Table~\ref{tbl:Z2_summary} in comparison with those from the XIS.

The 16--50 keV HXD periodograms, before and after the demodulation, 
are shown in Fig.~\ref{fig6:PG_nodem_dem}\,(b) 
in  black and red, respectively.
The demodulation parameters employed in calculating the red periodogram
are given in the figure.
Thus, by correcting the data  for the phase modulation 
with  $T=43.5$ ks  (Equation~\ref{eq:T_HXD}),
the HXD pulses, which were undetectable in the raw data,
have been clearly restored with $Z_4^2=40.64$,
at a period of 
\begin{equation}
P_{\rm HXD}= 5.209\:88 \pm 0.000\:03~{\rm s}.
\label{eq:P_HXD}
\end{equation}
The error is estimated in the same way as in Equation~(\ref{eq:T_XIS}).

The HXD is a low-background but non-imaging instrument,
and we are analyzing its data  without subtracting the background
which amounts to $\approx 91\%$ of the 16--50 keV events.
Therefore, we must examine 
whether the pulse-phase modulation is an artifact caused by background variations.
In the first 1/3 of the present observation,
the spacecraft was in such orbits  as to pass through the South Atlantic Anomaly,
where the background is higher and more variable than in the rest \citep{HXD2}.
We hence divided the HXD data into three disjoint time portions
with comparable durations,
and applied the demodulation analysis to them individually,
with the modulation  period fixed at $T=43.5$ ks  because it cannot be constrained.
Then, the three portions gave 
$(A, \psi)$=(1.0 s, 213$^\circ$), 
(1.4 s, 207$^\circ$), and  (1.1 s, 216$^\circ$) in this order,
together with $P$ which agrees with Equation~(\ref{eq:P_HXD}).
The parameters are thus similar among the three time portions, 
without correlation to the background behavior.
Considering further the XIS vs. HXD similarities in $T$ and $A$,
we conclude that the HXD results are not much
affected  by the background variations.
We also infer that the 43.5 ks phase modulation of the  HXD pulse is a coherent phenomenon,
because the three portions,  each covering about one modulation cycle,
indicate consistent values of $\psi$.

As given in Appendix A,  
the chance probability $Q_{\rm HXD}$,
for a value of $Z_4^2 \geq 40.64$
to arise via  statistical fluctuations,
is estimated as $Q_{\rm HXD} \approx 4\%$.
Since it is considered reasonably low,
the phase modulation is  likely to be real,
rather than due to statistical fluctuations.
The selection of $m=4$ is examined and justified in Appendix B.

For reference, we tried expanding the search range of $T$
up to 200 ks, considering the long data length (242 ks) of \NuS.
However, no additional DeMD peaks were found.

\subsubsection{Puzzles with the \Su\ data}
\label{subsubsec:demod_Suzazku_inconsistencies}
As seen so far, 
the XIS and HXD data  suggest 
that the HXC pulses of \sgr\ on this occasion were phase-modulated 
with $T=40-44$ ks and $A \approx 1.2$ s.
However, several inconsistencies and puzzles still remain within the HXD data,
and between the XIS and HXD data.
They are;

\begin{enumerate}
\setlength{\itemsep}{0mm}
\setlength{\itemindent}{-1mm}
\item[S1:]
When the lower energy bound of the HXD data is lowered from 16 keV,
the $T\approx 43.5$ ks DeMD peak  gradually diminishes,
down to $Z_4^2=33.09$ in 12--50 keV.
\item[S2:] 
The $T=43.5$ ks peak of the 16--50 keV HXD data 
is accompanied by $\psi=216^\circ$ (Table~\ref{tbl:Z2_summary}),
but it changes to $\psi\approx 60^\circ$
if using, {\em e.g.},  the 12--18 keV band instead.
The latter is not consistent,  either, with that from the XIS ($\psi\approx 160^\circ$).
\item[S3:]
The values of $T \approx 40.0$ ks indicated by the XIS (Equation~\ref{eq:T_XIS}) 
and $T \approx 43.5$ ks by  the HXD (Equation~\ref{eq:T_HXD})
appear somewhat discrepant.
\item[S4:]
As in Fig.~\ref{fig6:PG_nodem_dem}\,(b),
the optimum pulse period from the demodulated HXD data
falls on the lowest end of the conservatively estimated error range of $P_{\rm XIS}$.
\end{enumerate}

Among the above issues, [S1]  must be taken most seriously,
because it is specific to the HXD data,
and hence is free from the limited XIS time resolution.
It  is on one hand consistent with the pulse non-detection (using $m=2$)
in Fig.~\ref{fig3:PG_Suzaku_nodem}\,(b).
On the other hand, it is puzzling,
because expanding the energy range from 16--50 keV to 12--50 keV
should normally increase $Z_4^2$
(see an argument in the next subsection).
Therefore, we infer that the pulse coherence degrades
when a broader energy range is used,
as seen in Paper I.
We return to this issue after analyzing the \NuS\ data.

\begin{figure}[hbt]
\centerline{
\includegraphics[width=7.2cm]{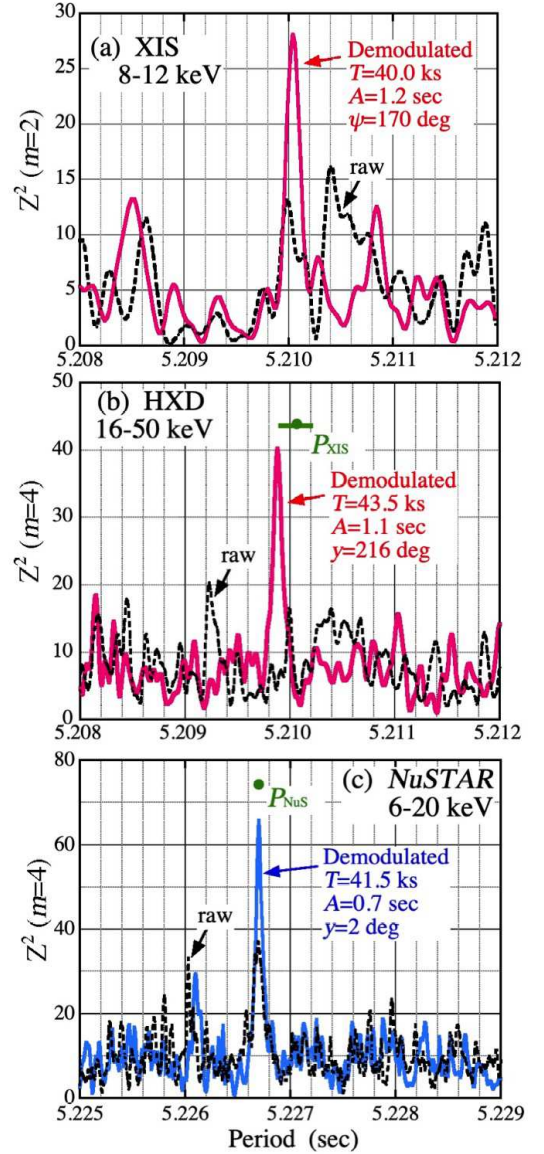}
}
\caption{Detailed pulse periodograms.
(a) The  \Su\ XIS result in 8--12 keV, where the dashed black curve denoted ``raw" 
uses the background-inclusive XIS photons without timing corrections,
whereas red is  that after applying the demodulation via Equation~(\ref{eq:modulation})
assuming $T=40.0$ ks.
The employed  $A$ and $\psi$ are given in the figure.
(b) The HXD result in 16--50 keV,
where the meanings of the curves are the same as  in panel (a).
The red trace employs $T=43.5$ ks.
The horizontal green  bar represents
$P_{\rm XIS}$  (with the associated errors).
(c)  Te 6--20 eV \NuS\ result, before (black) and 
after (blue) the demodulation with $T=40.5$ ks.
}
\label{fig6:PG_nodem_dem}
\end{figure}

\subsubsection{\NuS\ data}
\label{susubsec:demod_NuSTAR}
 We applied the same demodulation analysis to the \NuS\ data,
and obtained  the DeMDs in  panels (b1) and (b2) of  Fig.~\ref{fig5:Tscan_4results},
together with the periodograms in Fig.~\ref{fig6:PG_nodem_dem}\,(c).
The lower energy bound of 6.0 keV is chosen to approximately 
coincide with the SXC vs. HXC cross-over energy \citep{Tamba19},
and is made somewhat lower than that for the XIS (8 keV),
because the \NuS\ effective area at these energies decreases towards lower energies,
whereas that of the XIS behaves in the opposite sense.
The 6.0--20 keV DeMD reveals a strong peak with 
$Z_4^2=66.51$ ($\delta Z_4^2 =29.35$) at 
\begin{equation}
T = 40.5 \pm 0.8~~{\rm ks}~.
\label{eq:T_NuSTAR}
\end{equation}
The error is about half that in Equation~(\ref{eq:T_HXD}),
reflecting the \NuS\ data length which is about twice longer than that with \Su.
The corresponding DeMD peak is also recognized 
in the 20--60 keV result at a consistent $T$,
although it is considerably less conspicuous, 
and the  amplitude, $A \approx 1.2$ s, is somewhat 
larger than that in 6--20 keV, $A \approx 0.7$ s.

The reality of the above results was examined in several ways.
First, like in the HXD case, we applied the same analysis  
to the 1st and 2nd halves  of the  6--20 keV \NuS\ data,
this time allowing $T$ also to vary.
Then, the two halves yielded fully consistent DeMDs,
in terms of $T$, $A$, and $\psi$.
Next, like in \S~\ref{subsec:PG_nodem}, we changed $R_{\rm acc}$,
to find that the DeMD peak again becomes highest at $R_{\rm acc}\approx 50''$,
while the modulation parameters ($T$, $A$, and $\psi$) depend little
on $R_{\rm acc}$, at least from $35''$ to $100''$.
Therefore, the result is not likely an artifact
caused by the stray light from GRS 1919+105.
Finally, as given in Appendix A, 
the peak with $Z_4^2=66.51$ has a 
post-trial chance probability of
$Q_{\rm NuS} \approx 1\%$,
which is even lower than $Q_{\rm HXD}$.
We hence conclude that the phase modulation 
in the 6--20 keV  \NuS\ data is real.

\subsubsection{Puzzles with the \NuS\ data}
\label{susubsec:puzzles_NuSTAR}

From these DeMDs, we presume that the \Su\ and \NuS\ data 
recorded the same phenomenon.
The value of  $T$ indicated by  \NuS\ is in fact
consistent  with that of the XIS ($\approx 40.0$ ks; Equation~\ref{eq:T_XIS}),
However, $T$ could be inconsistent between
the HXD ($\approx 43.5$ ks; Equation~\ref{eq:T_HXD}) 
and \NuS\ ($\approx 40.5$ ks).
If so, the problem [S3] in \S~\ref{subsubsec:demod_Suzazku_inconsistencies}
is unlikely to be an artifact due to the limited time resolution of the XIS,
and must be regarded as inherent to the HXD data.

Even ignoring for the moment 
this HXD vs. XIS (plus \NuS) discrepancy in $T$,
the \NuS\ data themselves are subject to the following two puzzles.
\begin{enumerate}
\setlength{\itemsep}{0mm}
\setlength{\itemindent}{0mm}
\item[N1:]
When the two \NuS\ energy ranges used in Fig.~\ref{fig5:Tscan_4results},
6--20 and 20--60 keV, are added together, 
the 40 ks  DeMD peak decreases to  $Z_4^2 = 44.97$,
which is higher than that in 20--60 keV (33.47)
but much lower than in 6--20 keV (66.51).
\item[N2:]
As in Table~\ref{tbl:Z2_summary},
we find $\psi =3^\circ$ and $\psi =246^\circ$,
in 6--20 and 20--60 keV, respectively.
The latter,  calculated for  $T=38.7$ ks,
becomes  $\psi \approx 160^\circ$ if using $T=40.5$ ks.
Therefore,  the initial phase is  $\sim 180^\circ$ off
between the two energy ranges.
\end{enumerate}

Evidently, [N1] and [N2] are of the same nature 
as [S1] and [S2], respectively.
In particular, [N1] (as well as [S1])  is puzzling,
because periodic signals
with a constant pulse profile
and insignificant background 
should satisfy a relation as (Paper I) 
\begin{equation}
Z_m^2 \propto N_{\rm tot}\times ({\rm PF})^2
\label{eq:Z2_vs_PF}
\end{equation}
where $N_{\rm tot}$ is the total number of signal photons,
and  PF is the pulsed fraction.

\subsection{Pulse profiles}
\label{subsec:Pr_nodem_dem}

The issues [S1] and [N1] suggest that the pulse profiles and/or phases
are considerably energy dependent,
so the PF degrades when the energy range is expanded.
To examine this possibility,
let us look at  the folded pulse profiles
in various energies from the two data sets.

\begin{figure}[thb]
\centerline{
\includegraphics[width=8.7cm]{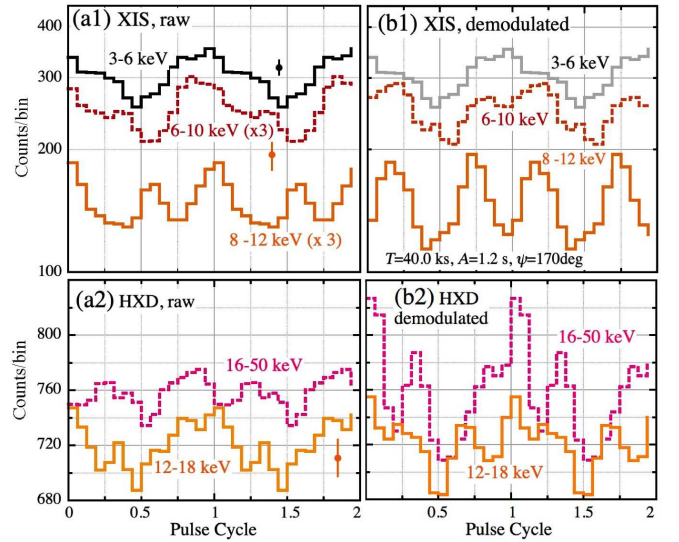}
}
\caption{Background-inclusive pulse profiles of SGR 1900+14 obtained with \Su,
shown for two cycles
 after applying the running average (see ext).
The ordinate is in units of  counts per bin.
(a1) Three-band XIS pulse profiles,
folded at $P_{\rm XIS}$ of Equation~(\ref{eq:P0_XIS}).
The ordinate is logarithmic,
and the 6--10 keV and 8--12 keV profiles  
are both multiplied by a factor of 3.
(a2) Results from the HXD in two energy intervals,
folded also at $P_{\rm XIS}$.
The ordinate is linear, because the data are background dominated.
(b1) The same as (a1), but the 6--10 keV and 8-12 keV results 
have been demodulated,
using $T=40.0$ ks, $A=1.2$ s, $\psi=160^\circ$, and $P=P_{\rm XIS}$.
(b2) The HXD profiles in the same energy bands as in (a2),
but demodulated  using $T=43.5$ ks, $A=1.1$ s, 
$\psi=225^\circ$, and $P=5.20987$ s. 
}
\label{fig7:Pr_Suzaku_nodem_dem}
\end{figure}

\break
Figure~\ref{fig7:Pr_Suzaku_nodem_dem} shows
background-inclusive pulse profiles
 with the \Su\ XIS (panel a1) and the HXD (panel a2),
folded under the conditions as specified in caption.
Here and hereafter, we include the XIS1 data,
because its background is stable on time scales of seconds.
The profile is always shown
after taking a running average,
where we smooth a  time series $\{x_j\}$
by replacing $x_j$ with $0.25 x_{j-1} + 0.5 x_j + 0.25 x_{j+1}$.
This reduces the statistical fluctuation
in each data bin to 0.61 times the original Poisson value (Paper I).
The derived profiles are single-peaked at $\lesssim 10$ keV,
and  changes into a double-peaked shape 
in $\gtrsim 10$ keV.
However, the profiles appear rather different
between the 8--12 keV XIS data and the 12--18 keV HXD data.

Through  demodulation using the parameters given in the caption,
the  profiles changed into as in panels (b1) and (b2).
The 6--10 and 8--12 keV profiles became double-peaked,
and that in 16--50 keV changed drastically; 
the PF increased, and several sharp features  emerged.
However, the 12--18 keV HXD profile is 
still different from the  8--12 keV XIS result.
In addition,  the deep pulse minimum appears to move,
in complex ways, across  the whole XIS plus HXD energy range.
These results support our view 
that  the issue [S1] stems from energy-dependent pulse-profile changes 
that cannot be rectified by the demodulation.

\begin{figure}[tb]
\centerline{
\includegraphics[width=8.8cm]{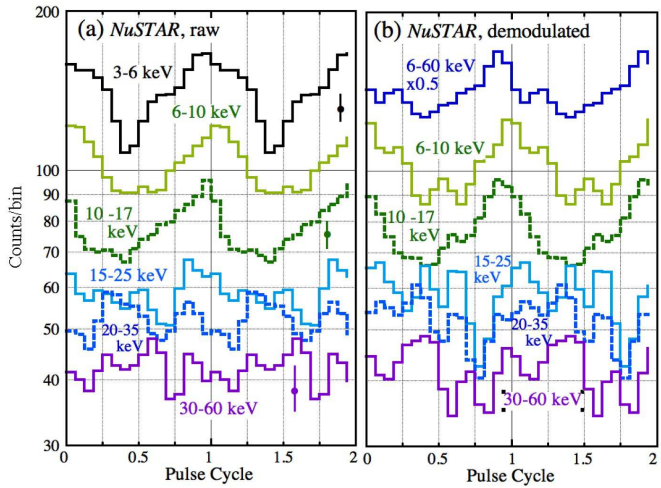}
}
\caption{Same as Fig.~\ref{fig7:Pr_Suzaku_nodem_dem},
but obtained with \NuS.
The ordinate is logarithmic.
(a) Pulse profiles in slightly-overlapping six energy intervals,
all folded at $P_{\rm NuS}$.
(b) The same as panel (a), but the profiles above 6 keV
have all been demodulated,
assuming  $T=40.5$ ks, $A=0.7$ s, 
$\psi=3^\circ$, and $P_{\rm NuS}$,
referring to Table~\ref{tbl:Z2_summary}.
The 3--6 keV profile in (a) is replaced with 
the demodulated 6--60 keV one (blue),
which is halved for presentation.
}
\label{fig8:Pr_NuSTAR_nodem_dem}
\end{figure}

The  background-inclusive \NuS\ pulse profiles 
are given in Fig.~\ref{fig8:Pr_NuSTAR_nodem_dem}\,(a),
again incorporating the running average.
Like in the \Su\ case,
the profiles are single-peaked up to the 10--17 keV range,
and become double- (or multiple-) peaked in 15--25 keV and beyond.
In panel (b), we applied the demodulation 
to the profiles except in  the lowest 3--6 keV range,
using a common set of  parameters 
determined in 6--20 keV (Table~\ref{tbl:Z2_summary}).
The  pulse amplitude increased mainly in $>15$ keV.
However, the pulse-phase assignment still remains ambiguous 
between  energies above and below $\sim  15$ keV.
In relation to the problem [N1],
this suggests considerable changes in the 
pulse properties across $\sim  15$ keV.
Furthermore, in both panels the pulse phase appears 
to advance from the 6--10 keV to 10--17 keV intervals.

With  both  \Su\ and \NuS, we have thus  confirmed 
that the demodulation actually increases the pulse significance,
and hence the PF from  Equation~(\ref{eq:Z2_vs_PF}),
at least in some energy intervals.
However, the pulse profiles still depend on the energy
in rather complex ways in both data sets;
presumably, this is responsible for  [S1] and [N1],
and possibly for [S2] and [N2] as well.
In addition, we have come across yet another issue
that is common to the two data sets:
\begin{itemize}
\setlength{\itemsep}{0mm}
\setlength{\itemindent}{1mm}
\item[SN1:]
The pulse-peak phase shifts gradually as a function of energy,
although the sign of this shift is not necessarily clear.
\end{itemize}

  \begin{table*}[hbt]
 \caption{A summary of the demodulation analysis, with 1-sigma errors.}
 \renewcommand{\arraystretch}{0.95}
 \begin{tabular}{lccccccc}
 \hline \hline 
 Energy (keV)& Condition$^{a}$  &  $P$    &  $Z_4^2$ & ${\delta Z_4^2}^{\,b}$ & $T$  & $A^{\rm c}$  & $\psi^{\,\rm d}$\\
 and $m$   &            &  (sec)   &                &                         &  (ks) &  (sec)  & (deg)     \\
 \hline \hline  
 \multicolumn{3}{l}{{\it Suzaku}}\\
 \hline  
8--12 (XIS 0+3)   &    Raw     &  $5.21000(15)$    & 12.29~   &  ---     &  ---      & ---      & ---    \\
 ($m=2$)        &Simple Dem. & $5.21004(12)$  & 26.36~  & 14.07  & $ 40.0^{+3.3}_{-2.5}$  & 1.3 & 160   \\
 \hline  
16--50  (HXD) &    Raw     &5.20999$^{c}~~~~$ &  16.57   &  ---      &  ---                     & ---      & ---    \\
($m=4$)     &Simple Dem. & 5.20987(3)~~ &  40.64   & 24.07  &  $43.5 \pm 1.8$  & 1.1  &  216    \\
 \hline  \hline
 12--50  (HXD)  &    Raw     &5.20984$^{e}~~~~$ & 17.21  &  ---       &  ---      & ---      & ---       \\
($m=4$)     &Simple Dem. & 5.20990(3)~~~& 33.09&  15.88 &  $43.7\pm 2.6$  & 1.2   &225   \\ 
                    &EDPV1           & 5.21002 (3)~~& 41.57 &   24.36 & $41.9 \pm 1.8 $&  1.1 &  153 \\ 
                    &EDPV2          & 5.21003 (3)~~& 47.98 &   30.77 & $41.2 \pm 1.2 $&  1.1 &  159 \\ 
  \hline  \hline
 \multicolumn{3}{l}{{\it NuSTAR}}\\
\hline  
6--20    &    Raw     &5.22669 (2) & 37.16     &  ---         &  ---      & ---      & ---     \\
$(m=4)$&Simple Dem. & 5.22671 (2) & 66.51    &29.35 &   $40.5\pm 0.8$  & 0.7   & 3  \\
\hline 
20--60   &    Raw       & 5.22669 (2) & 10.65  &  ---         &  ---      & ---      & ---     \\
$(m=4)$ &Simple Dem. & 5.22671 (2) & 33.47    & 22.82    &   $38.7^{+2.2}_{-1.0}$  & 1.2   & 246   \\
\hline 
6--60     &    Raw     &5.22672 (2) &  23.11  &  ---        &  ---      & ---      & ---    \\
$(m=4)$ &Simple Dem. & 5.22670 (2)  & 44.97  & 21.86   & $40.6\pm 1.1$ & 0.7  & 0  \\
              &EDPV1         & 5.22670 (2)  & 59.04  &  35.93   & $40.5\pm 0.7$ & 0.9  & 0 \\
               &EDPV2         & 5.22670 (2)  & 64.70  &  41.59   & $40.6\pm 0.5$ & 1.0  & 6  \\ 
\hline 
 \end{tabular}
 \label{tbl:Z2_summary}
 \smallskip
 \begin{itemize} 
 \setlength{\baselineskip}{2mm}
  \item[$^{\rm a}$]: ``Raw"= without timing correction, 
``Simple Dem."=with energy-independent demodulation via Equation~(\ref{eq:modulation}), ``EDPV1"=using Equation~(\ref{eq:EDPV1}), and 
 ``EDPV2"=incorporating equations~(\ref{eq:EDPV1}) and (\ref{eq:S(E)}).
  \item[$^{\rm b}$]: Increment in $Z_4^2$  from the ``Raw" value.
 \item[$^{\rm c}$]:  The errors associated with $A$ are  typically $\pm 0.3$ for \Su\
 and $\pm 0.2$ for \NuS.
 \item[$^{\rm d}$]:  The errors associated with $\psi$ are  typically $\pm 30^\circ$ for \Su\ 
 and $\pm 15^\circ$ for \NuS, reflecting the overall data length.
  \item[$^{\rm e}$]: The error is not estimated because the peak is not outstanding.
   \end{itemize}
  \end{table*}  

\subsection{Analysis of the soft-component signals}
\label{subsec:Tscan_SXC}

To  examine whether the SXC signals also suffer the pulse-phase modulation,
we applied the same  analysis 
to the 1--5 keV XIS data (combining the three cameras) with $m=2$,
and the 3--5 keV \NuS\ data with $m=4$.
The upper energy bound,  5 keV,
was selected  to exclude the HXC,
and the choice of $m$ is the same as before.
The inclusion of the XIS1 data is
because its background is negligible at these energies.

The derived soft-band DeMDs are presented in Fig.~\ref{fig9:Tscan_SXC}.
Although a broad enhancement is seen in the XIS DeMD (panel a)
over a range of  $T= 40-45$ ks,
the increment is only $\delta Z_2^2 \sim 4$, 
and the associated amplitude, $A\sim 0.2$ s,
is only $\sim 4\%$ of $ P$.
We do not find particular $Z_4^2$  enhancements
at $T\sim 40$ ks of the \NuS\ DeMD (panel b), either,
even though the allowed values of $A$ are relatively large,
 $A \lesssim 0.8$ s,
due to the small photon number (1658 events) in this energy band.
More generally, around the mean of $\langle Z_4^2 \rangle=41.98$
in the \NuS\ DeMD,
$Z_4^2$ is seen to fluctuate by $\pm3.63$ (1$\sigma$),
in agreement with the expectation of
 $\sqrt{4m}=4.0$. 

We hence conclude that the SXC pulsation is free  from the phase modulation 
that affects the HXC pulses.  This agrees  with the results
on the preceding two magnetars
(\citealt{Makishima16, Makishima19}; Paper~I),
and indicates a basic difference
between the two spectral components
in their timing properties.

\begin{figure}[bth]
\centerline{
\includegraphics[width=7.7cm]{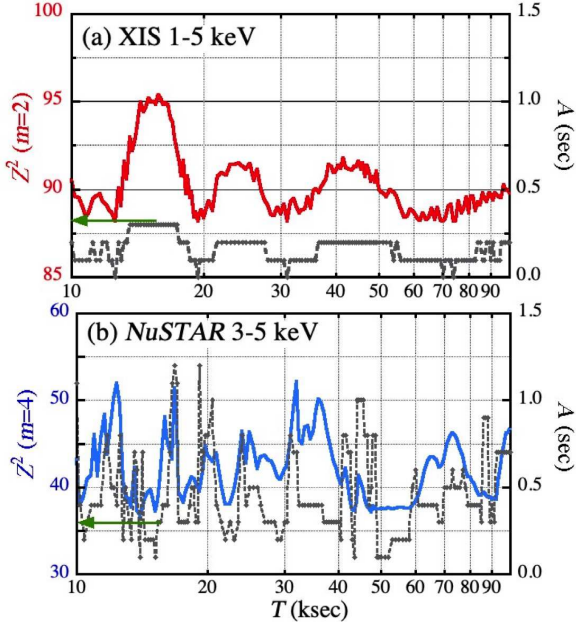}
}
\caption{Soft-band DeMDs, obtained with 
the \Su\ XIS in 1--5 keV using $m=2$ (panel a),
and with \NuS\ in 3--5 keV using $m=4$ (panel b).
The meanings of red/blue and dashed gray traces are the same
as in Fig.~\ref{fig5:Tscan_4results}.
}
\label{fig9:Tscan_SXC}
\end{figure}

\break
\section{ADVANCED TIMING STUDIES}
\label{sec:2nd_stage_ana}
Although the demodulation analysis was partially successful,
we are still left with the problems:
[S1] --[S4] (\S~\ref{subsubsec:demod_Suzazku_inconsistencies}),
[N1], [N2](\S~\ref{susubsec:puzzles_NuSTAR}),
and [SN1] (\S~\ref{subsec:Pr_nodem_dem}).
We suppose that these issues at least partially arise via
energy dependences  in  the pulse-phase-modulation phenomenon (Paper I).
By empirically  modeling these effects,
we hope to solve or explain the issues,
in terms of the basic dynamics of an axial rigid body,
{\em i.e.,}  rotation around the symmetry axis
and free precession.

\subsection{Preliminary evaluations}
\label{subsec:Wdrop_estimates}

A possibility suggested by  [S1], [S2], [N1],  and [N2]
is that $\psi$ in Equation~(\ref{eq:modulation})  
depends on the energy,
changing considerably across a narrow interval around 15 keV.
Actually, such an effect was observed from \oneE\ (Paper I);
as the energy increases from 10 keV to 27 keV, 
$\psi$ decreased by $\sim 65^\circ$,
followed by a $180^\circ$ jump.

To  examine the above possibility,
we conducted a few  preliminary tests.
Figure~\ref{fig10:Wdrop_estimates} (a1) and (a2) show 
so-called  double-folded maps using the \Su\ HXD data.
The abscissa is the pulse phase $\Phi/2\pi$,
the ordinate (from top to bottom) 
the modulation phase $\Psi/2\pi$, 
and the colors represent the photon intensity.
The value of $\Psi$ at the observation start  is 
$\psi$ in Equation~(\ref{eq:modulation}).
As in  Paper I, the running average is 
applied  in the $\Phi$  dimension, but not in $\Psi$.
The correction for exposure is applied only in $\Psi$,
because it is highly uniform in $\Phi$.
In both panels, the pulse peak forms a yellow vertical ridge at  $\Phi/2\pi  \approx 1.0$,
but  it wiggles as a function of $\Psi$,
just visualizing  the pulse-phase modulation.
The lateral swing of the ridge in  (a2), $\approx \pm 0.2 P$, 
agrees with the observed $A\approx1.0$ s.
Moreover, the wiggles  in the two panels occur in the  opposite sense;
in (a1), it is most delayed  in $\Phi$
at $\Psi/2\pi \approx 0.3$ and most advanced at $\Psi/2\pi\approx 0.7$,
but in (a2) the largest delay and advance 
occur at  $\Psi/2\pi \approx 0.9$ and  $ \approx 0.4$, respectively.
This confirms that $\psi$ changes 
by $\approx 180^\circ$ between the two energy bands.

Using the  \NuS\  data, we conducted a more quantitative evaluation,
to obtain  the results in Fig.~\ref{fig10:Wdrop_estimates}\,(b).
In  several energy bands (with partial overlaps),
we calculated $Z_4^2$ as a function of $\psi$,
keeping $T=40.5$ ks but allowing $P$  and $A$  to vary.
In the 6--12 keV band, $Z_4^2$ became highest at $\psi \sim 0^\circ$,
but  toward higher energies this peak  increased in $\psi$,
and reached $\psi \approx 160^\circ$ at the 20--60 keV range.
This is consistent with the implications from \Su,
and supports our conjecture,
although $\psi$ {\em increases} towards higher energies
contrary to the behavior of \oneE.

In the \NuS\ observation of \oneE\  in 2016,
not only $\psi$ but  also  $A$
exhibited strong energy dependences,
with  a factor $\sim 3$ enhancement
at $22 \pm 7$ keV (Paper I).
In the present case, such behavior is not observed,
because we always find $A \approx 1.0$ s within $\pm 30\%$
(Table~\ref{tbl:Z2_summary}).
We hence treat $A$ as an energy-independent constant,
although we do not require it to be the same between the two observations.

\begin{figure}[hbt]
\centerline{
\includegraphics[width=8.3cm]{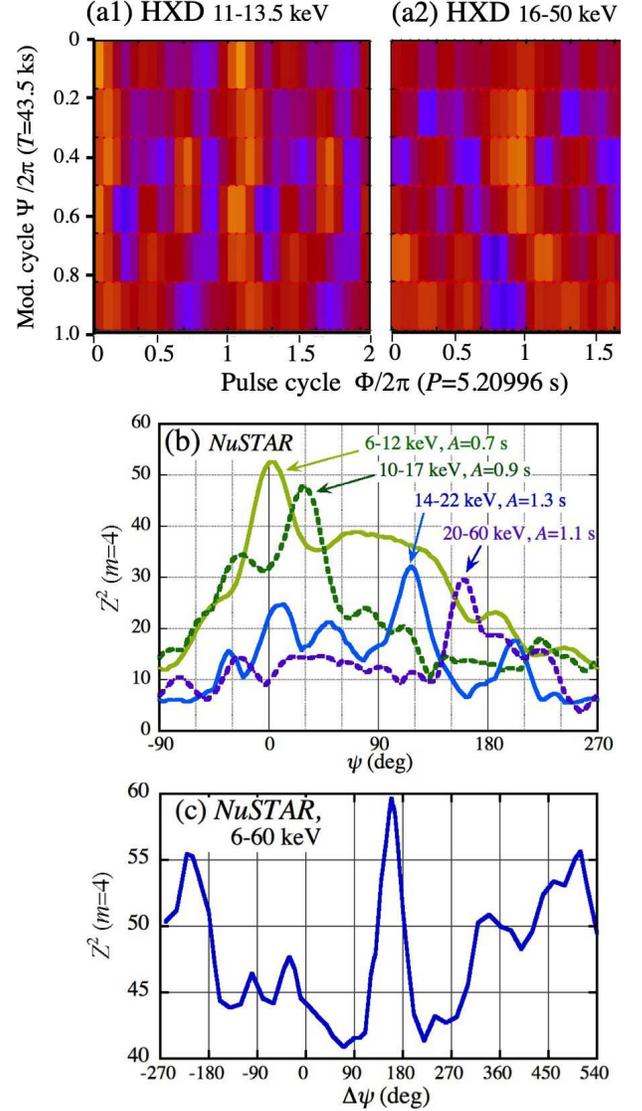}
}
\caption{(a1) Double-folded map (see text) produced from the 11.0--13.5 keV HXD data
using $P$ of Equation~(\ref{eq:P_HXD})
and $T=43.5$ ks of Equation~(\ref{eq:T_HXD}),
where the photon intensity is represented by the color gradient
(yellow, red, and blue from brightest to faintest).
(a2) Same as (a1), but using the 16.0--50.0 keV HXD data.
(b) Maximum values of $Z_4^2$ 
in four energy intervals of  the \NuS\ data,
shown as a function of $\psi$. 
(c) The highest $Z_4^2$ from the 6--60 keV \NuS\ data,
derived via the EDPV1 correction, shown as a function of $\Delta \psi$.
Except $T=40.5$ ks, all the  parameters are free.
}
\label{fig10:Wdrop_estimates}
\end{figure}

\subsection{Formalism}
\label{subsec:EDPV_formalism}

Following Paper I, we modify Equation~(\ref{eq:modulation}),
and empirically model the HXC pulse timing behavior as
\begin{equation}
\delta t =  P \cdot S(E)/360+ A \sin \left[ 2\pi t/T - \psi(E) \right] ~,
\label{eq:EDPV2}
\end{equation}
which we  call {\it Energy Dependent Pulse-phase Variation} (EDPV).
Here, $E$ is  the energy in units of keV,
and $\psi(E)$, in place of  the constant $\psi$ in Equation~(\ref{eq:modulation}),
describes how the modulation phase varies with  $E$.
The other variable $S(E)$ describes, in units of degree (0 to $360^\circ$),
the energy dependence of the pulse phase  [SN1].
On a double-folded map like Fig.~\ref{fig10:Wdrop_estimates}\,(a1,a2),
$\psi(E)$  and $S(E)$  specify vertical  ($\Psi$ direction) 
and horizontal  ($\Phi$ direction) displacements of the pulse pattern,
respectively, both as functions of $E$.
While  $\psi(E)$ is coupled with the pulse-phase modulation at $T$,
$S(E)$ is not.

Based on the preliminary evaluations, as well as Paper I,
let us model $\psi(E)$  as 
\begin{equation}
\psi(E) = 
\left\{
\begin{array}{ll}
\psi_0                                                     & (E \leq E_1)\\
\psi_0 + \Delta \psi \; (E-E_1) /( E_2-E_1)    & (E_1 < E < E_2 )\\
\psi_{\rm 0}+ \Delta \psi                           & (E_2 <E)
\end{array}
\right.
\label{eq:EDPV1}
\end{equation}
where $E_1$, $E_2$, and $\Delta \psi$ are adjustable parameters,
and  $\psi_0$ gives the initial modulation phase at  $E\leq E_1$.
Thus, $\psi(E)$ remains at $\psi_0$ for $E \leq E_1$,
changes linearly by $\Delta \psi$ from $E_1$ to $E2$,
and stays at $\psi_0+\Delta \psi$ for $E \geq E_2$.
(Compared to Paper I, $\Delta \Psi$ is here defined using the opposite sign,
to make the present results easier to grasp.)
This modeling  is hereafter referred to  as EDPV1.

We model $S(E)$ in a  parabolic way as (Paper I)
\begin{equation}
S(E) = 
\left\{
\begin{array}{ll}
0                                                     & (E \leq 8~{\rm keV})\\
\frac{R}{(E_3-8)} (E-8)(E_3-E)   & (E>8~{\rm keV})~,\\
\end{array}
\right.
\label{eq:S(E)}
\end{equation}
using two parameters $R$ and $E_3$.
Thus,  $S(E)$ is assumed to work in $E>  8$ keV 
(see \S~\ref{subsubsec:EDPV_NuSTAR}),
with a slope $R \equiv (dS/dE)_{\rm 8keV}$  at 8 keV
in units of deg keV$^{-1}$.
If $R>0$, $S(E)$ increases up to $E=(8+E_3)/2$ keV
where it reaches $R\, (E_3-8)/4$,
then it starts decreasing, returns to 0 at $E_3$,
and becomes negative for $E>E_3$.
If $R<0$, $S(E)$  behaves in the opposite way.
The timing correction by  Equation~(\ref{eq:S(E)}),
together with Equation~(\ref{eq:EDPV1}),
is hereafter called EDPV2 modeling.

Below, we focus on the HXD and   \NuS\  data
in the 12--50 keV and 6--60 keV bands, respectively,
which are nearly the widest ranges 
where the HXC is clearly detected.
The  EDPV1 scheme is first applied to the data,
to confirm its effectiveness,
and to optimize its parameters ($\Delta \psi$, $E_1$, and $E_2$),
separately for the two data sets.
Then, we proceed to the EDPV2 corrections.
These attempts are not performed on the XIS data.
 
  \begin{table*}
 \caption{Summary of the optimum EDPV parameters, with 1-sigma errors.}
 \label{tbl:EDPV_summary}
 \begin{tabular}{lccccccccc}
 \hline \hline 
  Function      & \multicolumn{4}{c}{$\psi(E)$} && \multicolumn{2}{c}{$S(E)$}  \\
    \cmidrule{2-5} \cmidrule{7-8}
  Parameer   &  $\Delta \psi $& $\psi_0^{\,a,b}$ &  $E_1$  & $E_2$ &&$R$           & $E_3$  & $T^{\, a}$  & $A^{\, a}$  \\
                       &       (deg)          & (deg)       &  (keV)     & (keV)    && (deg/keV) & (keV)\\
  \hline 
\multicolumn{5}{l}{\Su\ HXD (12--50 keV)}  \\
~~~EDPV1    &$184\pm 20 $   & 153  &$13.5\pm 0.3 $ & $15.8 \pm 0.3$ &&---                    &---              & 41.9& 1.1\\
~~~EDPV2    &$174\pm 22$   & 159  &$13.5\pm 0.4 $ & $15.6 \pm 0.5$ &&$-4.6 \pm 2.3$& $41 \pm 4$&41.2  & 1.1 \\
\hline 
\multicolumn{5}{l}{\NuS\  (6--60 keV)}\\
~~~EDPV1     &$162\pm 12$   &   0     &$13.3\pm 1.7 $ & $20.5 \pm 2.4$&&---                       &---            & 40.5  & 0.9\\
~~~EDPV2    &$160\pm 15$    &   6     &$13.2\pm 1.8 $ & $21.0 \pm 2.6$&&$1.3 \pm 1.2$  & $68^{+40}_{-24}$  & 40.6 & 1.1 \\
  \hline 
 \end{tabular}
  \begin{itemize} 
 \setlength{\baselineskip}{4mm}
  \item[$^{\rm a}$]: See Table~\ref{tbl:Z2_summary} for errors of these quantities.
  \item[$^{\rm b}$]: Between \Su\ and \NuS, $\psi_0$ can differ,
   because it is determined by the start time of each observation.
  \end{itemize}
 \end{table*}

\subsection{Results}
\label{subsec:EDPV_results}

\subsubsection{The 12--50 keV HXD data}
\label{subsubsec:EDPV_HXD}

Substituting  $\psi(E)$ into Equation~(\ref{eq:EDPV2})
and  setting $S(E)=0$,
we applied the EDMP1 correction
to  the 12--50 keV HXD data.
Starting from initial guesses of $E_1=14.0$ keV, $E_2=16.0$ keV,
and $\Delta \psi =180^\circ$,
as suggested by [S2] and  Fig.~\ref{fig10:Wdrop_estimates}\,(a1,a2),
we trimmed these parameters (as well as $P$, $A$, and $\psi_0$),
so as to maximize the DeMD peak.
Then, $Z_4^2$ has increased  (Table~\ref{tbl:Z2_summary}),
and  yielded the  parameters as in Table~\ref{tbl:EDPV_summary}.
In contrast, assuming $\Delta \psi \sim -180^\circ$
did not  increase $Z_4^2$.
Therefore, like in  the \NuS\ case (Fig.~\ref{fig10:Wdrop_estimates}\,b),
$\psi(E)$  is thought  to {\em inrease} by  $\sim 180^\circ$
from  13.5 keV to 15.8 keV.

The 12--50 keV HXD data  were further analyzed via the EDPV2 scheme,
by activating $S(E)$.
We optimized $R$ and $E_3$, 
as well as the EDPV1 parameters which changed to some extent.
The EDPV1/2 parameters determined in this way are 
also given inTable~\ref{tbl:EDPV_summary}.
The DeMD peak further increased,
and $R$ turned out to be negative
(see a later discussion in \S~\ref{subsubsec:Pr_EDPV}).

Figure~\ref{fig11:Tscan_EDPV}\,(a) superposes
DeMDs from the 12--50 keV HXD data, 
derived under three conditions;
via  the simple demodulation (dashed black),
the EDPV1 modeling (orange), 
and  the  EDPV2  scheme (red).
Their basic features  are summarized in Table~\ref{tbl:Z2_summary}.
The  progressively more complex timing corrections
have  produced the following four noticeable effects.
\begin{enumerate}
\setlength{\itemindent}{0mm}
\item The DeMD peak became higher,
from $Z_4^2 = 33.09$ to $Z_4^2 = 41.57$ (EDPV1),
and further to $Z_4^2 = 47.98$ (EDPV2).
Thus, [S1] was mostly solved.
\item From the 12--50 keV HXD data,
the EDPV1 correction deduced $\psi_0=153^\circ$,
which agrees with $\psi= 160^\circ$ (Table~\ref{tbl:Z2_summary})
of  the 8-12 keV XIS data.
The EDPV2 scheme further enhanced the agreement.
This means that [S2] was mitigated.
\item The peak centroid of the HXD DeMD evolved 
from  $T\approx 43.5$ ks  to  $T \approx 42$ ks (EDPV1),
and finally to $T \sim 41$ ks (EDPV2)
which is fully consistent with  those from the XIS and \NuS.
Therefore, [S3] was solved rather unexpectedly,
although its mechanism is unclear.
\item The best pulse period  with the HXD
was at  first described by Equation~(\ref{eq:P_HXD}),
but it has increased  to 5.21003 s,
which agrees well with $P_{\rm XIS}$.
Therefore,  [S4] was also solved automatically.
\end{enumerate}

To elucidate the item 4 above,
Fig.~\ref{fig12:EDPV2_results}\,(a) compares
four pulse periodograms, all from the 12--50 keV HXD data
but derived in four different ways as explained in the caption.
It visualizes  that the  series of timing corrections
have not only increased the pulse significance,
but also brought the HXD pulse period
in a full agreement with $P_{\rm XIS}$, thus solving  [S4].

\begin{figure}[hbt]
\bigskip
\centerline{
\includegraphics[width=8.2cm]{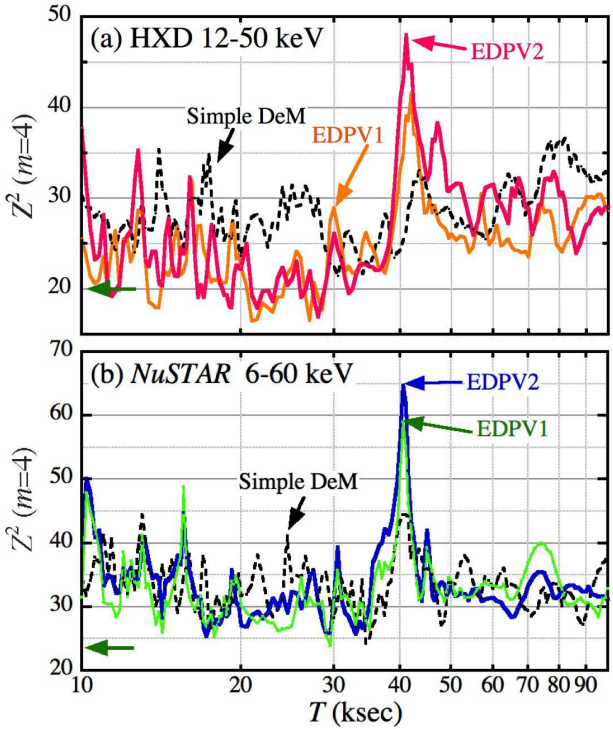}
}
\caption{DeMDs ($m=4$) derived under three different conditions,
from the 12--50 keV HXD data (panel a)
and the 6--60 keV \NuS\ data (panel b).
The result of the simple demodulation with Equation~(\ref{eq:modulation})
is in dashed black,
that with the EDPV1 correction is in orange or light green, 
and the EDPV2 result is in red or blue.
As $T$ is varied, $P$, $A$, and $\psi_0$ are allowed to vary,
whereas the other EDPV$i$ ($i=1$ or 2) parameters are 
fixed to those in Table~\ref{tbl:EDPV_summary}, 
separately for the HXD and \NuS.
}
\label{fig11:Tscan_EDPV}
\end{figure}

\subsubsection{The 6--60 keV \NuS\ data}
\label{subsubsec:EDPV_NuSTAR}

We applied the same EDPV1 modeling
to the 6--60 keV  \NuS\ data.
Figure~\ref{fig10:Wdrop_estimates}\,(c) depicts
how the pulse significance varied with $\Delta \psi$,
when $T=40.5$ ks is fixed but all the other parameters 
($P$, $A$, $\psi_0$, $E_1$, and $E_2$) are allowed to vary.
The data  give a clear constraint as $\Delta \psi =162^\circ \pm 15^\circ$,
where the error has been determined in the same way as before.
The result also reveals  a pair of subsidiary peaks
which are  $\pm 180^\circ$ off the central peak,
but they are lower by $\approx 5$ in heights,
so that their occurrence probability is
each $\sim 1/6$ of that of the central peak.
This difference can be explained in the following way:
if we select $\Delta \psi$ from either side peak,
the coherence in $\Psi$ between the $E \leq E_1$ and $E \geq E_2$ regions
becomes the same as the case with $\Delta \psi=162^\circ$,
but the coherence in $\Psi$ must be lost between $E_1$ and $E_2$,
causing a decrease  in $Z_4^2$.
We hence adopt the central peak,
in agreement with Fig.~\ref{fig10:Wdrop_estimates}\,(b) 
and our conclusion from the HXD data.
Including this $\Delta \psi$, the EDPV1 parameters from \NuS\ are
given in  Table~\ref{tbl:EDPV_summary}.

We analyzed the 6--60 \NuS\ data 
further employing the EDPV2 corrections,
and  obtained the optimum parameters as 
given also  in Table~\ref{tbl:EDPV_summary}.
Compared with the HXD results, 
$\Delta \psi$ and $E_1$ can be regarded as the same within errors,
whereas $E_2$ and $E_3$ are  higher. i 
Unlike the HXD case,  $R$ became marginally positive,
and its absolute value is considerably smaller.
Therefore, some of the EDPV2 parameters are considered to change with time,
just as  $A$ varies on time scales of months to years \citep{Makishima19}.

\begin{figure*}[bth]
\centerline{
\includegraphics[width=11cm]{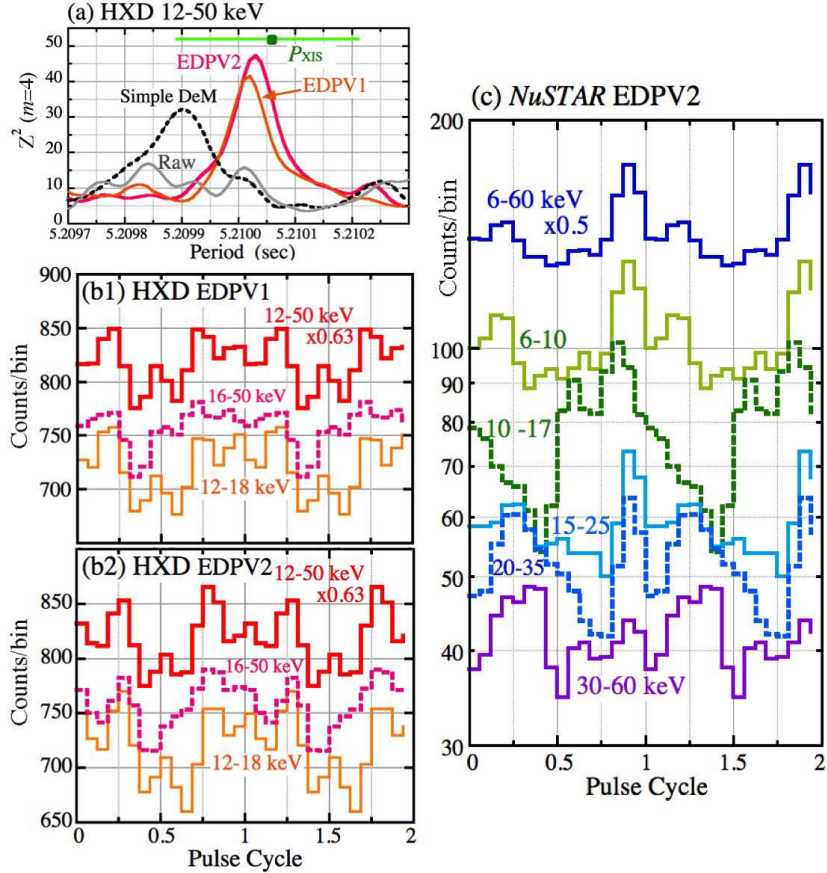}
}
\caption{Results from the EDPV analysis.
(a) Pulse periodograms with the 12--50 keV HXD data,
computed under 4 conditions;
without timing corrections (``Raw," gray),
with the simple demodulation (dashed black),
the EDPV1 modeling (orange),
and the  EDPV2 correction (red).
(b1) HXD pulse profiles with the EDPV1 scheme.
(b2) Same as (b1) but with the EDPV2 correction.
(c) \NuS\ pulse profiles (originally in Fig.~\ref{fig8:Pr_NuSTAR_nodem_dem})
after the EDPV2 processing.}
\label{fig12:EDPV2_results}
\end{figure*}

Figure~\ref{fig11:Tscan_EDPV}\,(b) presents
the DeMDs from the 6--60 keV \NuS\  data,
calculated under the same three conditions as for the HXD data.
Their basic properties are again given in Table~\ref{tbl:Z2_summary}.
The DeMD peak has become significantly higher,
from $Z_4^2=44.97$ with the simple demodulation (dashed black)  
to $Z_4^2=59.04$ (EDPV1; light green),
and further  to $Z_4^2=67.40$ (EDPV2; blue),
though still smaller than that ($Z_4^2=66.51$)
derived in 6--20 keV via the simple demodulation.
The final value of $\psi_0 =6^\circ$ agrees
with $\psi =3^\circ$ obtained originally in 6--20 keV.
In the course of these improvements,
the best values of $T$, $A$, and $P$ have remained relatively unchanged.
Thus, the EDPV2 modeling on the \NuS\ data
has solved [N2], and at least partially [N1] as well.

So far, we used the start energy of 8 keV in the $S(E)$ formalism
(equation~\ref{eq:S(E)}),
but this is somewhat arbitrarily.
We hence  tried changing it to 6 keV or 10 keV,
to find that  the case with 8 keV is slightly preferred by the \NuS\ data. 
(The 12--50 keV HXD data are almost insensitive.)
This  justifies  our use of 8 keV as the start energy,
both for \Su\ and \NuS.

\subsubsection{Pulse profiles}
\label{subsubsec:Pr_EDPV}

Figure~\ref{fig12:EDPV2_results}\,(b1) shows
the folded pulse profiles from the HXD data,
processed through the EDPV1  scheme.
Compared to the results with the simple demodulation 
in Fig.~\ref{fig7:Pr_Suzaku_nodem_dem}\,(b2),
the 16--50 keV pulse amplitude became smaller for some reasons,
but the profiles became  less energy dependent.
They exhibit  a relatively symmetric pair of horn-like peaks 
at $\Phi/2\pi  \approx \pm 0.3$,
which are similar to those found in the  XIS profile
in 8--12 keV (and to a lesser extent in 6--10 keV).
The HXD profiles also show another pair of weaker 
peaks  at  $\Phi/2\pi  \approx -0.1$ and $\approx +0.5$,
and the four peaks are spaced by  $\approx 1/4$ cycles.
Furthermore,  the 12--18 keV profile is 
seen to lag behind that in 16--50 keV.
This ``soft lag" presumably demanded  the negative value of $R$.
(Due to the insufficient XIS time resolution,
we cannot draw any conclusion about  XIS vs. HXD time lags.)

Figure~\ref{fig12:EDPV2_results}\,(b2) shows 
the HXD profiles corrected for the EDPV2 effect,
using the parameters in Table~\ref{tbl:EDPV_summary}
which dictate
that the HXD pulse phase is most advanced at $E=(8+E_3)/2 = 25.5$ keV, 
by $\Delta \Phi = R\, (E_3-8)/4 =-38^\circ$ ($-0.11$ pulse cycles).
As a result, the profiles have become mostly free from the soft lag,
and exhibit the four-peak structure  more clearly than before.
As an assuring fact, 
such a four-peak profile was observed from \sgr\  
during its Giant Flare in 1988 \citep{Feroci01},
and from 4U~0142+61 when demodulated \citep{Makishima14}.
On the other hand, this raises a suspicion
that we might be biased towards particular EDPV2 solutions
that selectively enhance the $m=4$ power.
So, in Appendix B,
we repeated the analysis by changing $m$,
and removed this concern.

The \NuS\ pulse profiles, 
originally in Fig.~\ref{fig8:Pr_NuSTAR_nodem_dem},
became as in Fig.~\ref{fig12:EDPV2_results}\,(c),
after processed through the EDPV2 modeling.
(We skip showing the EDPV1 results, because they are rather similar.)
The profiles (in particular that in 10--17 keV) still depend on the energy,
but they commonly exhibit a narrow peak at $\Phi/2\pi \approx -0.1$
which are nearly in phase across the entire entire energy.
On both sides of this main peak,
we observe secondary peaks at a separation 
 by 1/3 to 1/4 cycles, just like in the demodulated HXD profiles
 (see also Appendix B).

\subsubsection{Significance of the EDPV effects}
\label{subsubsec:EDPV_significance}

The increase in the pulse significance,
from the simple demodulation to the EDPV1 modeling,
is $\delta Z_4^2=8.48$ and $14.07$,
with \Su\  (12--50 keV) and  \NuS\ (6--60 keV), respectively.
The  implied decrease in $Q$ is 
a factor of $2.7 \times 10^{-2}$ (\Su)  and $1.9 \times 10^{-3}$ (\NuS).
Since these values are not too small,
we may not readily exclude the chance origin of these improvements,
when  the increase in the  parameters ($\Delta \psi$, $E_1$, and $E_2$)
are considered.
Unlike the case of the simple demodulation,
it is not easy, either, to conduct any simulation studies,
using the actual or Monte-Carlo-simulated data.

In spite of these limitations,
we regard the EDPV1 improvements 
(\Su\ and \NuS\ altogether) as real, for several reasons.
First,  the energy-dependent changes in $\psi$ are directly visible from the data  
(\S~\ref{subsec:Wdrop_estimates}; Fig.~\ref{fig10:Wdrop_estimates}).
Second,  the EDPV1 corrections have not only increased $Z_4^2$,
but also solved (at least partially)  the puzzles [S1]-[S4], [N1], and [N2].
Third,  the \Su\ and \NuS\ solutions agree within errors
on $\Delta \psi$ and $E_1$, even though they differ in $E_2$.
Such a coincidence would not easily happen 
if the pulse enhancements were  simply due to chance fluctuations.
Finally, a very similar phenomenon has already 
been confirmed in \oneE\ (Paper I)
at broadly similar energies, from 10 to 30 keV.

The case for the EDPV2 corrections,
where we further incorporate  $S(E)$, is more subtle.
Probably it is not very significant for \NuS,
because $\Delta Z_4^2 = 5.66$ from EDPV1 to EDPV2 is rather small,
in agreement with the fact that $R=0$ is marginally excluded.
Actually, the apparent soft lag seen in Fig.~\ref{fig8:Pr_NuSTAR_nodem_dem},
between the 6--10  and 10--17 keV profiles,
was mostly rectified  (though not shown) by the EDPV1 corrections,
and  a minor hard lag remained, which demanded $R>0$ in EDPV2.
In contrast,  on the HXD data,
the EDPV2 scheme (with $\Delta Z_4^2 = 6.41$ over the previous step)
is considered more effective,
because it removed the 12--18 keV versus 16--15 keV soft lag
(Fig.~\ref{fig12:EDPV2_results})
which was left over by the EDPV1 step.
This agree with the positive value of $R$ 
in the EDPV2 solution.
Moreover, the 12--50 keV profile has become sharper
by the EDPV2 step.

\section{DISCUSSION}
\label{sec:discussion}

\subsection{Summary and evaluation of the results}
\label{subsec:summary}

We analyzed the \Su\ and \NuS\ data of \sgr,
acquired in 2009 and 2016, respectively.
Through the epoch-folding analysis incorporating $Z_m^2$ statistics,
the source pulsation in the SXC was clearly detected,
with the \Su\ XIS at $P_{\rm XIS}$ (Equation~\ref{eq:P0_XIS}),
and with \NuS\ at $P_{\rm NuS}$ (Equation~\ref{eq:P0_NuSTAR}).
In both data sets, the SXC pulses were quite regular,
without evidence  for any periodic phase modulation  (\S~\ref{subsec:Tscan_SXC}).

The HXC pulses, which were not detected via simple epoch-folding analysis
either with the  \Su\ HXD or \NuS\ (\S~\ref{subsec:PG_nodem}; 
Fig.~\ref{fig3:PG_Suzaku_nodem}\,b; Fig.~\ref{fig4:PG_NuSTAR_nodem}\,c),
have been detected significantly by both  these instruments
through the demodulation correction
(\S~\ref{subsec:DeMDs}; Fig.~\ref{fig5:Tscan_4results};  Fig.~\ref{fig6:PG_nodem_dem}).
Specifically, the 6--20 keV \NuS\ data revealed a prominent $Z_4^2$ 
increase at $T=40.5$ ks  (Equation~\ref{eq:T_NuSTAR}) 
with a chance probability of $Q_{\rm NuS} = 1\%$
(\S~\ref{susubsec:demod_NuSTAR}),
and  the 8--12 keV XIS data on the HXC 
indicated  a consistent  $T$ (\S~\ref{subsubsec:demod_XIS}).
The DeMD with the 16--50 keV HXD data also revealed a clear  peak,
with $Q_{\rm HXD} = 4\%$ (\S~\ref{subsubsec:demod_HXD}),

If we were allowed to regard
the HXD and \NuS\ results as the same phenomenon,
the overall chance probability of our finding would be
\begin{equation}
Q_{\rm tot}= Q_{\rm HXD} \times Q_{\rm NuS} = 4 \times 10^{-4}
\label{eq:total_probability}
\end{equation}
which is extremely low.
However, the HXD-indicated $T=43.5$ ks (Equation~\ref{eq:T_HXD})
is somewhat inconsistent with
those from the XIS and \NuS\  [S3] (Table~\ref{tbl:Z2_summary}).
Further inconsistencies were  found 
between the XIS and HXD  [S4],  
within the HXD  data [S1,S2],
and within those of  \NuS\ [N1,N2].
The energy-dependent pulse-phase shift
was identified as  yet another issue [SN1].
Thus, the simple energy-independent pulse demodulation
was only partially successful in recovering  the HXC pulses,
so Equation~(\ref{eq:total_probability}) 
needs some reservations.

Assuming that the HXC pulses are subject to EDPV effects,
we attempted further arrival-time corrections
(\S~\ref{subsec:EDPV_formalism}),
employing  two functions $\psi(E)$ and $S(E)$,
which describe the pulse-pattern shifts 
in the $\Psi$ and $\Phi$ directions, respectively.
Using the 12--50 keV HXD data  and the 6--60 keV  \NuS\ data 
as  fiducial energy ranges,
and guided by preliminary studies (\S~\ref{subsec:Wdrop_estimates}),
we identified the EDPV1+2 parameters 
(separately for \Su\ and \NuS; Table~\ref{tbl:EDPV_summary})
that maximize the pulse significance,
and solved the puzzles [S1], [S2], [N1], and [N2].
The EDPV1+2 corrections also brought 
the XIS and HXD pulse periods into an agreement,
identified $T= 40.5$ ks as a common periodicity, 
and  aligned up the pulse profiles from either data set 
throughout the energy.
Thus,  [S3],  [S4], and [SN1] have been solved.

In the above analyses, the EDPV1+2 parameters have been 
determined solely to maximize $Z_4^2$ for the HXC signals
in the respective fiducial energies;
no attempt was made to bring $T$,
the most fundamental parameter,
in agreement among the different instruments.
Nevertheless, the agreement on $T$ has been achieved automatically.
Therefore, the two observations are considered 
to have witnessed the same phenomenon.
Then, we are allowed to quote Equation~(\ref{eq:total_probability})
in its face value, 
and conclude that the HXC pulse-phase modulation is real,
rather than due to statistical fluctuations.

These results have the following meanings, 
which affirmatively answer the three objectives described in \S~\ref{sec:intro}.
\begin{enumerate}
\setlength{\itemsep}{1mm}
\item
The presence of a significant pulse-phase modulation in the HXC,
and its absent in the SXC,
agree with the behavior of  4U~0142+61 and \oneE.
Thus, \sgr\ provides a third example to show this behavior.
Our conjecture, that the phenomenon should be 
detected from nearly all  magnetars,
was reinforced.
\item
The pulse-phase modulation of the HXC
exhibits significant energy dependences,
{\em i.e}., the changes in $\psi$ by $\Delta \psi \sim 180^\circ$
across  an energy interval from $\approx13.5$ keV to 15.8-21.0 keV.
These effects, first seen in  \oneE\ (Paper I),
are hence suggested to be not rare among magnetars.
\item
The \Su\ and \NuS\  results agree on 
essential features of the phenomenon,
including its energy dependence.
This mitigates the risk of instrument-specific artifacts.
They however differ  
in some  details ({\em e.g.}, on $E_2$ and $R$);
the phenomenon is hence considered time variable to some degree.
\end{enumerate}

\subsection{Interpretations of the Results}
\label{subsec:EDPV_interpretation}

\subsubsection{Dynamics of an axially-symmetric rigid body}
\label{subsubsec:rigidbody_dynamics}

As developed through a series of our studies,
the phase modulation of the HXC pulses,
which is now confirmed in \sgr,
can be described using  the basic dynamics of an axisymmetric rigid body.
Namely, we identify $T=40.5$ ks 
with the slip period of the NS in \sgr,
which is axially elongated by $\epsilon \approx  P/T = 1.3 \times10^{-4}$
and performs free precession.
The deformation can be ascribed to the 
stress by extreme $B_{\rm t}$ which reaches $\sim 10^{16}$ G.

To be more concrete,
consider two Cartesian frames with a common origin (Fig.18 of \citealt{Makishima21});
an inertial frame $\Sigma =(\hat X,\hat Y, \hat Z)$
with the unit vector $\hat Z$ parallel to ${\bf L}$, 
and $\Sigma^* =(\hat{x}_1, \hat{x}_2, \hat{x}_3)$ fixed to the NS.
As before, $\hat{x}_3$ is identified with the NS's symmetry axis.
The  triplet $(\Phi, \alpha, \Psi)$ provides the three  Euler angles,
which specify the instantaneous attitude of $\Sigma^*$
relative to $\Sigma$.
While  $\alpha$ is constant, $\Phi$ and $\Psi$ both vary.
Every time $\Phi$ completes its one cycle  
(in the period $P_{\rm pr}$, or one pulse), $\Psi$ advances by 
$2 \pi\epsilon \cos \alpha=2\pi (P_{\rm pr}/T)$ (\S~\ref{sec:intro}).
Assuming $\cos \alpha \approx 1$,
$\Psi$ returns to its initial value in the slip (or beat) period $T$,
which comprises $T/P_{\rm pr}$ precession cycles and
$(T/P_{\rm pr})+1$ rotations around $\hat{x}_3$.

We further assume that the X-ray emission pattern at each energy is constant
when described in the $\Sigma^*$ coordinates, 
and the changes of $\Sigma^*$  relative to $\Sigma$
are responsible for all observed variability, 
including the pulsation and its phase modulation. 
When the emission pattern is symmetric around $\hat{x}_3$,
and hence independent of $\Psi$,
the pulsation will be strictly periodic,
like what we observe  for the SXC.
If instead the emission breaks symmetry around $\hat{x}_3$ (and hence in $\Psi$),
the pulse-peak phase becomes dependent on $\Psi$,
as modeled by Equation~(\ref{eq:modulation}) in the simplest case,
and actually  exhibited by the HXC from \sgr.

In short, the pulses from a rotating NS become phase-modulated
when the following three symmetry breakings all take place;
(i) $\alpha \ne 0$, (ii) $\epsilon \ne 0$,
and (iii) an asymmetric emission pattern around $\hat{x}_3$.
The  HXC of the relevant  magnetars is thought to
satisfy all these conditions, 
whereas the SXC only (i) and (ii).
These clear distinctions  between the SXC and HXC,
in their timing behavior and their spectral shapes,
suggest a fundamental difference in their origins.

\subsubsection{A possible geometry}
\label{subsubsec:possible_geometry}

The above general conditions can be satisfied by
a specific  geometry given in Paper I,
which also affords an explanation of  the EDPV effects.
Suppose that the SXC is emitted by a region symmetric around $\hat{x}_z$,
whereas the HXC, arising via, {\em e.g.}, the two-photon process,
has a conical beam pattern around $\hat{x}_z$.
The cone brightness is assumed to vary with $\Psi$ (the cone azimuth)
due, {\em e.g}., to  the presence of local multipoles 
which break the symmetry around $\hat{x}_z$.
Then, the pulse-phase modulation up to $A \sim P/4$ 
can be explained in a semi-quatitative way (Paper I).
Furthermore, let  the directional vector $\hat \xi$ 
represent the generatrix along which the HXC is brightest.
If $\hat \xi$   moves in $\Psi$  with energy,
due to some strong-field physics 
such as proton cyclotron resonances (Paper I),
the EDPV1 effects can be explained.

As mentioned in Paper I,
the EDPV2 effect represented by $S(E)$ is more difficult to interpret.
To see this, let $\Pi_3$ be the plane defined by 
$\hat{Z} \| {\bf L}$ and $\hat{x}_3$,
which rotates around $\hat{Z}$ with a period $P_{\rm pr}$.
Then, $\hat \xi$ as defined above,
rotates relative to $\Pi_3$ with a period $T$, 
in which it crosses $\Pi_3$ twice.
At every crossing, the pulse arrival-time delay $\delta t$ will change its sign.
When averaged over a cycle,
we expect $\langle \delta t \rangle \approx 0$, or $S(E) \approx 0$,
as long as Equation~(\ref{eq:modulation}) provides  a good approximation.
However, if the time profile of $\delta t$ is much deviated from a sinusoid,
and  asymmetric between $\delta t>0$ and $\delta t<0$,
we may expect $S(E) \ne 0$.
Thus, we tentatively regard $S(E)$ as a {\em modeling artifact},
which would vanish when we improve Equation~(\ref{eq:modulation}).
Such attempts were already made in Paper I,
but only very preliminarily.

\subsubsection{Implications for the nature of magnetars}
\label{subsubsec:magnetar_nature}
So far, we have adopted the  interpretation of mangetars
as isolated NS powered by magnetic energies \citep{Mereghetti08}.
Some alternative models however describe them 
as NSs with ordinary dipole magnetic fields  as $B_{\rm d} \sim 10^{12}$ G,
powered by mass accretion from fossil or fallback 
disks around them  \citep[e.g.][]{accretion}.
In fact, infrared observations provided evidence for such disks 
around some magnetars, including  4U~0142+61 in particular
\citep[e.g.][]{disk}.

Based on the disk-accretion scenario, \cite{Grimani21} argued 
that the 55 ks modulation in 4U 0142+61 can be explained 
when the source is hidden periodically by the disk
if it is in a Keplerian rotation and is free-precessing.
However, the disk is so distant (Appendix C)
that  the emission region would look point-like,
and the disk is not  highly ionized.
Then, the modulation must get stronger towards lower energies
because of the increasing photo-absorption,
contradicting to the general absence 
of phase modulation in the SXC pulses.
Therefore, this scenario would work for neither 4U~0142+61,
nor \sgr\ which is in a similar condition.
Some other mechanisms must be sought for
if the observed phenomenon it to be explained 
by the disk-accretion scenario.

Regardless of details of the phase-shift production,
the EDPV effects  in the two objects
must be explained.
Taking  \sgr\ for example, 
the 12 keV pulses at a particular phase in $\Psi$
need to arrive by $\sim 1$ s {\em earlier}  than  expected,
whereas the 20 keV pulses {\em later}  by a similar amount
(Fig.~\ref{fig10:Wdrop_estimates}). 
Such a sharp  energy dependence is incompatible with
the {\em broadband} nature of  X-rays from accretion columns,
wherein  the $\sim 12$ keV and $\sim 20$ keV photons  
must behave in  positively correlated ways.
In contrast, our scenario based on the strong-field physics
(\S~\ref{subsubsec:possible_geometry})
can explain the essential timing properties of \sgr\ including its EDPV behavior.
Therefore, the X-rays produced via accretion, if any,
should contribute little to the overall X-ray emission.

We also examined how  a circum-stellar disk 
affects the NS dynamics by {\em forced precession}.
This may take place through two channels;
one is via  the accretion torque,
wherein the NS can be spherical but needs to be accreting.
The other  is via direct gravity;
the NS needs to be deformed,
but the accretion is not required.
As given in Appendix C,
the former mechanism predicts a long period of
forced precession as  $\sim 2\times 10^3$ yr,
whereas the latter is even longer
by many orders of magnitude.
Thus, the rigid-body dynamics of these magnetars are 
not affected by the  disks around them.

Although we have  argued  against the accretion scenario,
we do not mean that NSs with $B_{\rm d}\gtrsim 10^{14}$ G
cannot become accreting sources, or cannot reside in binaries.
In fact, the binary X-ray pulsar X-Persei,
accreting from the companion's stellar winds,
has been found to have  $B_{\rm d} \sim 1 \times 10^{14}$ G \citep{Yatabe18}.
As another interesting case,
the gamma-ray binary LS~5039
is likely to harbor a non-accreting magnetar,
whose magnetic energy is  released via  
interactions with the primary's stellar winds
to drive the remarkable non-thermal activity 
\citep{Yoneda20, Yoneda21}.
Yet another example is so-called Central Compact Objects (CCOs),
rather inactive NSs found at the center of some supernova remnants
\citep[e.g.][]{CCO}.
They are thought to have weak $B_{\rm d}$  but intense $B_{\rm t}$, 
and the latter sustains their activity.
Thus, a fair fraction of NSs may be born as magnetars in a broad sense \citep{Nakano15}, 
and reside in various environments.
As suggested by the present study,
some, if not all, of them might have $B_{\rm t} \sim 10^{16}$ G.
It would hence be an interesting future work
to classify magnetized  NSson the $(B_{\rm d}, B_{\rm t})$ plane.
}

\subsection{\bxp\ as a counter example}
\label{subsec:4U1626}

The confirmed phase-modulation period $T$ is 
rather similar among the three objects;
55 ks in 4U~0142+61, 36 ks in \oneE, and 40.5 ks in \sgr.
Then, a concern arises;
could this effect be some instrumental or observational artifacts  in hard X-rays,
and  emerge virtually in all X-ray pulsars?
As a candidate counter example,
we  studied  the ultra-compact binary pulsar \bxp,
because its pulse period, 7.68 s, is similar to those of magnetars,
and its orbital Doppler effect is very small as 
$<13$ lt-ms  \citep{4U1626Doppler}.

On  2006 March 9 through 11 (ObsID 40015010),
\bxp\ was observed with \Su\ for an elapsed time of 239 ks.
The data were already analyzed by \citet{Iwakiri12},
who detected the pulsation at $P=7.67795(9)$ s.
We processed these HXD data  in the same way as for \sgr,
including the barycentric correction,
and  applied the demodulation analysis to
the 12--34 keV and 34--40 keV events.
The former energy range enables us to utilize the highest data statistics,
whereas the latter  to emulate the  pulse significance actually observed from \sgr.

\begin{figure}[t]
\centerline{
\includegraphics[width=7.2cm]{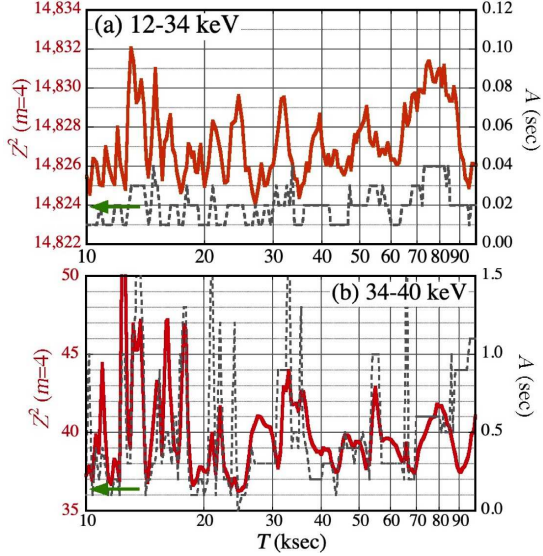}
}
\caption{DeMDs ($m=4$) from the \Su\ HXD data of \bxp,
in the (a) 12--34  keV and (b) 34--40 keV  intervals.
The plot style is the same as for Fig.~\ref{fig5:Tscan_4results}.
}
\label{fig13:4U1626}
\end{figure}

The  DeMDs from  \bxp\ are shown in Fig.~\ref{fig13:4U1626}.
In  panel (a) for 12--34 keV, 
$Z_4^2$ takes extremely large values
with the  mean of $\langle Z_4^2 \rangle=14,827$,
reflecting high signal statistics.
Nevertheless,  the DeMD  does not show outstanding peaks,
and varies only by  $\pm 1.80$ (1$\sigma$)
which is  smaller than the Poissonian prediction 
(\S~\ref{subsec:Tscan_SXC}).
Moreover, the modulation amplitude  is  tightly constrained as 
$A \lesssim 0.04$ s ($\lesssim 0.5\%$ of $P$) throughout.

Similarly, the 34--40 keV DeMD in Fig.~\ref{fig13:4U1626}\,(b) 
has a 1$\sigma$ variation by $\pm 2.66$,
without significant peaks except in 12--18 ks
where multiples of the \Su's orbital period appear.
Here, $A$ takes much larger values than in panel (a),
for the following reason \citep{Makishima16}.
In general, increasing $A$ has two opposite effects;
to degrade the underlying pulse coherence,
and to increase the number of different combinations of the Poisson fluctuations.
When the pulsation has high significance,
the former effect dominates to favor small values of $A$,
whereas the opposite occurs 
when the underlying pulsation has  low statistics.

The two DeMDs in Fig.~\ref{fig13:4U1626} can  thus be
interpreted both as a sum of a regular pulsation
and Poissonian fluctuations,
with no evidence for intrinsic pulse-phase modulation
over the 10--100 ks interval.
This conclusion remains unaffected
even if using various different energy intervals,
or extending the search range of $T$ up to $\sim 500$ ks
(beyond which the data are no longer constraining).
Therefore, \bxp\ provides a good counter example to our concern,
and suggests that the pulse-phase modulation
is an effect specific to  the HXC of  magnetars,
rather than some observational artifacts.

In terms of the rigid-body dynamics,
the negative result on \bxp\ can be explained 
by logically inverting the three symmetry-breaking conditions
described in \S~\ref{subsubsec:rigidbody_dynamics},
and restoring the individual symmetry.
That is, the phase modulation will vanishes if
(\rm i$'$)  $\alpha \approx 0$ (aligned rotators), 
or (ii$'$)  $\epsilon \approx 0$ (spherical symmetry),
or (iii$'$) $\vec{\xi} \| \hat{x}_3$.
Among the three options,  (i$'$) is  ruled out because the source is pulsing.
Next, consider  (ii$'$).
Since \bxp\ has $B_{\rm d} \approx 3 \times 10^{12}$ G \citep{Iwakiri12},
it would be very likely to have $B_{\rm t} \ll 10^{16}$ G.
Then, the NS should be nearly spherical with  $\epsilon \ll 10^{-4}$,
satisfying (ii$'$). 
In this case, Equation (\ref{eq:slip_period}) indicates $T \gg 10^4 P_{\rm pr}$.
Then, during a typical observation,
$\vec{\xi}$  is  phase-locked to the $\Pi_3$ plane,
and the pulses will keep a constant phase.
Finally, let us consider (iii$'$).
The X-rays from accreting pulsars  are produced in their accretion columns
mainly via thermal Compntonization,
modified by electron cyclotron resonances.
So  the emission would be azimuthally isotropic,
even though otherwise in the polar direction.
Therefore,  (iii$'$) is also likely to hold.

To  summarize, the emission from accreting pulsars, 
including \bxp,
is in a condition (i)$\land$(ii$'$)$\land$(iii$'$),
where $\land$ means {\em logical and}.
In contrast, the SXC and HXC of magnetars are expressed 
by (i)$\land$(ii)$\land$(iii$'$) and  (i)$\land$(ii)$\land$(iii), respectively.
As a result, the pulse-phase modulation is observed
only from the HXC of magnetars.
These characterizations highlight the intrinsic difference
between the magnetically-powered and accretion-powered NSs.

\section{CONCLUSION}
\label{sec:conclusion}

\begin{enumerate}
\setlength{\itemsep}{0mm}
\setlength{\itemindent}{0mm}
\item
In hard X-rays,
the 5.2-s pulsation of 
SGR 1900+14 was not detected at first,
with either \Su\ or \NuS.
However, the pulses became detectable
by correcting the photon arrival times
for the phase-modulation effects,
assuming a period of $T=40.5 \pm 0.8$ ks
as consistently indicated  by the \Su\ XIS, 
\Su\ HXD, and  \NuS.
Thus, \sgr\ becomes a third example 
that shows this behavior, after 4U~0142+61 and \oneE.
\item
We identify $T$  with the slip period
associated with  free precession of the NS,
which is axially deformed to $\epsilon \approx  P/T =1.3 \times 10^{-4}$,
presumably by  the stress of $B_{\rm t} \sim 10^{16}$ G.
The observed value of $A \approx 1.0$ s is consistent with this picture.
\item
A series of problems, left by the simple demodulation,
were mostly solved by considering the EDPV effects,
like in the \NuS\ data of \oneE.
The derived EDPV parameters  partially agree
between \Su\ and \NuS.
\item
The pulse-phase modulation is not likely to be an observational artifact,
because it was absent in the \Su\ data of a counter example, \bxp.
\item This phenomenon is possibly  ubiquitous among magnetars, 
and will provide valuable clues to 
their $B_{\rm t}$,
as well as  to their HXC emission mechanism
which must be distinct from that of the SXC.
\item
The present results  favor the interpretation of magnetars
as magnetically powered NSs,
rather than as those accreting from circumstellar disks.
\item
Intense toroidal agnetic fields, up to $B_{\rm t} \sim 10^{16}$ G,
could be rather common among magnetars
and similar NSs.
\end{enumerate}

\break
\section*{Acknowledgements}
The present work was  supported in part by the JSPS
grant-in-aid (KAKENHI), number 18K03694.
The authors would like to  thank the anonymous referee
for constructive comments.

\section*{Appendix A: Statistical significance of the phase modulation}

By the simple energy-independent demodulation analysis
in \S~\ref{subsec:DeMDs},
we obtained $Z_4^2=40.64$ and $Z_4^2=66.51$,
from the 16--50 keV HXD data and the 6--20 keV \NuS\  data,
respectively (Table~\ref{tbl:Z2_summary}).
Referring to the chi-square distribution with $2m=8$ degrees of freedom,
the chance occurrence probability becomes
$Q_{\rm HXD}^0=2.4\times 10^{-6}$ (HXD)
and $Q_{\rm NuS}^0=2.4\times 10^{-11}$ (HXD).
Although the significance appears to differ by many orders 
of magnitude between the two results,
the difference is in reality  not so large,
because in terms of the increment,
the \NuS\ peak has $\delta Z_4^2=29.35$ 
whereas that of the HXD is $\delta Z_4^2 24.07$,
with a difference of only 5.28.

To evaluate the true statistical significance of the detected effect,
we must consider two additional factors \citep{Makishima16}.
One is that these values must be  multiplied by the total number 
of {\em independent} trials (difficult to estimate)
involved in the DeMDs in Fig.~\ref{fig5:Tscan_4results}.
The other is that the above estimates of $Q$ are valid 
only when the signal is purely Poissonian,
and needs revisions
when the signal is already pulsing before the demodulation.
Hence, we developed several methods
to estimate the true significance
(\cite{Makishima14, Makishima16}, Paper I),
mainly using the actual data 
but partially incorporating Monte-Carlo technique.

Adopting the method in Paper I,
 we  repeated the same demodulation analysis
over an interval of $T =0.05-3.0$ ks, 
which is shorter than the orbital period of \Su\
but still longer than the pulse period.
We varied the scan steps in  $T$ as  $\Delta T=T^2/T_{\rm tot}$,
where $T_{\rm tot} = 114$ ks is the total observation time lapse.
This $\Delta T$ is  the smallest step 
that ensures the independence between adjacent sampling points
in terms of Fourier wave numbers.
We have thus obtained $2242$ ($=114/0.05-114/3.0)$ steps in $T$, 
which is 219 times larger than that in the actual DeMD calculation
over  the $T=10.0-100$ ks interval,
namely, $114/10.0 - 114/100 = 10.26$.
We scanned $P$, $A$, and $\psi$ over the same ranges and same steps
as in deriving Fig.~\ref{fig5:Tscan_4results}\,(a2),
so that the trial numbers in these quantities are the same.
Through this {\em control} study,
$Z_4^2$ exceeded the target value  $Z_4^2 = 40.64$ at  8 values of $T$.
Therefore, the probability 
for the 43.5 ks peak in the 16--50 keV DeMD 
to appear by chance finally becomes
$
Q_{\rm HXD} = 8/219 \approx 4\%~.
$

We conducted the control study using the 6--20 keV \NuS\  data as well,
exactly in the same manner as for the HXD data,
but with $\Delta T$  halved because of the twice longer $t_{\rm tot}$.
The ratio of 219, 
between  the control study and the DeMD in Fig.~\ref{fig5:Tscan_4results}\,(b1),
remains the same.  (In calculating the latter, $T$ is over-sampled.)
As a result, the control study yielded two ponts in $T$
where $Z_4^2$ exceeds the target value of 66.51.
Therefore, the chance probability for the 40.5 ks peak
in the 6--20 keV \NuS\ DeMD is estimated as
$
Q_{\rm NuS} = 2/219 \approx  1\%~
$
which is lower, as expected, than $Q_{\rm HXD}$.

\medskip
\section*{Appendix B: Effects of the Fourier harmonic number}

To examine whether our results are biased by our choice of $m=4$,
we repeated the EDPV2 analysis  on the 12--50 keV HXD data, 
by changing $m$ from 1 to 8,
and re-optimizing the EDPV2 parameters at each $m$.
The obtained  values of $Z_m^2$ are given  Table~\ref{tbl:m-dependence},
together with the difference $\Theta_m \equiv Z_m^2 - Z_{m-1}^2$
which represents  the $m$-th Fourier power of the pulse profile.
For a purely Poissonian signal, we expect  $\Theta_m \approx 2$ regardless of $m$.
Thus, the highest power is in $m=4$, 
and then in $m=8$,
both in agreement with  the pulse profiles in Fig.~\ref{fig12:EDPV2_results}.

As for the 6--60 keV \NuS\ data, the same scan in $m$ gave
$\Theta_m=28.3, 4.6, 18.5$, and 12.4, for $m=1,2,3$, and 4, respectively.
Thus, the 4th harmonic is somewhat weaker than the 3rd.
Nevertheless, our choice of $m=4$ for the \NuS\ data is considered appropriate,
because we found $\Theta_m \lesssim 3$ for $m\geq 5$.

\begin{table}[h]
\caption{Maximum values of $Z_m^2$ from EDPV2 analysis of the 12--50 keV HXD data,
as a function of $m$ (see text).}
\label{tbl:m-dependence}
\vspace*{-2mm}
\begin{center}
\begin{tabular}{lcccccccc}
\hline \hline
$m$           &  1  & 2  &  3  &  4  &  5&  6  & 7 & 8\\
\hline
$Z_m^2$  &14.2  & 25.5 & 31.6 & 48.0 & 49.6 & 50.9 & 54.9 & 63.1 \\
\hline
$\Theta_m$ & 14.2 & 11.3 & 6.1 & 16.4 & 1.6 & 1.3 & 4.0 & 8.2\\
\hline
\end{tabular}
\end{center}
\end{table}

\section*{Appendix C: Forced precession induced by a circum-stellar disk}
Of the two modes of forced precession induced by a circum-stellar disk
(\S~\ref{subsubsec:magnetar_nature}),
the accretion-induced effect will take place on a time scale 
which is comparable to the spin-up time scale of the NS, $\tau_{\rm su}$.
According to \cite{GL79}, it is estimated as
\[
\tau_{\rm su} \sim 3 \times 10^4 P({\rm sec})^{-1} \mu_{30}^{-2/7}\, L_{36}^{-6/7} ~{\rm yr}
\]
where $\mu_{30}$ is the magnetic moment  of the NS in $10^{30}$ cgs units,
and $L_{36}$ is the accretion luminosity in units of $10^{36}$ erg s$^{-1}$.
Then, even if assuming  the  most favorable conditions
that $L_{36} \sim 0.55$ of \sgr\ is totally due to accretion,
and yet the NS has $B_{\rm d} \sim 10^{14}$ G  implying $\mu_{30} \sim 100$,
a rather long time scale as $\tau\sim 2 \times10^3$ yr is indicated.

The other mode, namely, the direct gravitational perturbation
on an axially deformed NS,
was  studied by \cite{forced_precession}.
In this case, the period of forced precession of the  NS  
is given as
\[
P_{\rm forced} =(16 \pi^2R^3) /(9G M_{\rm d} \epsilon P)
\]
(adapted from Equation 2 of \citet{forced_precession})
where $G$ is the gravitational constant,
$M_{\rm d}$ is the total disk mass,
and  $R$ is a representative disk radius.
For simplicity, we  assumed that the disk normal is parallel to $\bf{L}$.
If considering the disk around 4U~0142+61,
\cite{disk} give $M_{\rm d} \sim 6\times 10^{28}$ g 
and $R \sim 4 \times 10^{11}$ cm.
These, together with  $\epsilon \sim 10^{-4}$
from our measurements and $P=8.96$ s, 
yield $P_{\rm forced} \sim 10^9$ yr.
Therefore, the effect is totally negligible.
Although the estimate may change to some extent
when considering the disk wobbling,
 the conclusion would remain unaffected.

\end{document}